\documentclass[preprint,onecolumn,superscriptaddress,preprintnumbers]{revtex4}

\usepackage[utf8]{inputenc}
\usepackage{graphicx,comment}
\usepackage{amsmath}
\usepackage{amssymb}
\usepackage{hyperref}
\usepackage{cleveref}
\usepackage{mathrsfs}  
\usepackage[dvipsnames]{xcolor}
\usepackage[normalem]{ulem}
\usepackage{url}

\newcommand\be{\begin{equation}}
\newcommand\ee{\end{equation}}
\newcommand\bea{\begin{eqnarray}}
\newcommand\eea{\end{eqnarray}}

\usepackage{color}

\begin{document}

\title{Milli-Magnetic Monopole Dark Matter and the Survival of Galactic Magnetic Fields }
\author{Michael L. Graesser}
\email{michaelgraesser@gmail.com}
\affiliation{Theoretical Division, Los Alamos National Laboratory, Los Alamos, NM 87545, USA}
\author{Ian M. Shoemaker}
\email{shoemaker@vt.edu}
\affiliation{Center for Neutrino Physics, Department of Physics,Virginia Tech, Blacksburg, Virginia 24061, USA}
\author{Natalia Tapia Arellano}
\email{ntapiaa@vt.edu}
\affiliation{Center for Neutrino Physics, Department of Physics,Virginia Tech, Blacksburg, Virginia 24061, USA}

\date{\today}

\preprint{LA-UR-21-24525}

\begin{abstract} 

Dark sectors with Abelian gauge symmetries can interact with ordinary matter via kinetic mixing. In such scenarios, magnetic monopoles of a broken dark $U(1)$ will appear in our sector as {confined} milli-magnetically charged objects under ordinary electromagnetism. 
{Halo ellipticity constraints are shown to significantly bound
the strength of dark magnetic Coulomb monopole interactions.}
{The bound monopole ground state, which in vacuum is stable and has no magnetic charge or moment, is 
shown to become quantum mechanically unstable 
in the presence of an external, ordinary magnetic field.}
If these states contribute sizably to the local dark matter density, they can extract significant energy from the galactic magnetic field {if their decay occurs on a galactic timescale or less}.
We revise and extend this ``Parker Bound'' on galactic magnetic energy loss to milli-monopoles which leads to the strongest existing constraints on these states, {satisfying our halo ellipticity bounds}, over a wide range of monopole masses.  
\end{abstract}

\maketitle
\tableofcontents

\newpage

\section{Introduction}

Although it is known that the most abundant matter in the Universe is a non-luminous ``Dark Matter (DM),'' its particle nature is unknown. At present, it is not known if the DM sector constitutes one or many new fields,
or whether the DM itself is fundamental or composite.  A large class of models fall into the category of ``dark sector'' models in which the new DM states are uncharged under the Standard Model (SM) gauge symmetries. In the case that the dark sector contains a new dark $U(1)$ symmetry, one cannot forbid a gauge invariant and renormalizable term in the Lagrangian~\cite{Holdom:1985ag}
\be   
\mathscr{L} \supset \varepsilon F^{\mu\nu} F'_{\mu\nu},
\ee
where $\varepsilon$ is a dimensionless parameter controlling the strength of this ``kinetic mixing'' interaction, while $F^{\mu\nu}$ and $F'_{\mu\nu}$ are respectively the electromagnetic and dark field strengths. 
At large distances, electrically charged particles under electromagnetism appear millicharged under the $U(1)_{D}$.  A large amount of phenomenological work has focus on the consequences of ``electric'' charges in the dark sector (e.g.~\cite{Batell:2009di,Aguilar-Arevalo:2018wea,Arguelles:2019xgp}). 

\begin{figure}[t!]
\includegraphics[angle=0,width=.45\textwidth]{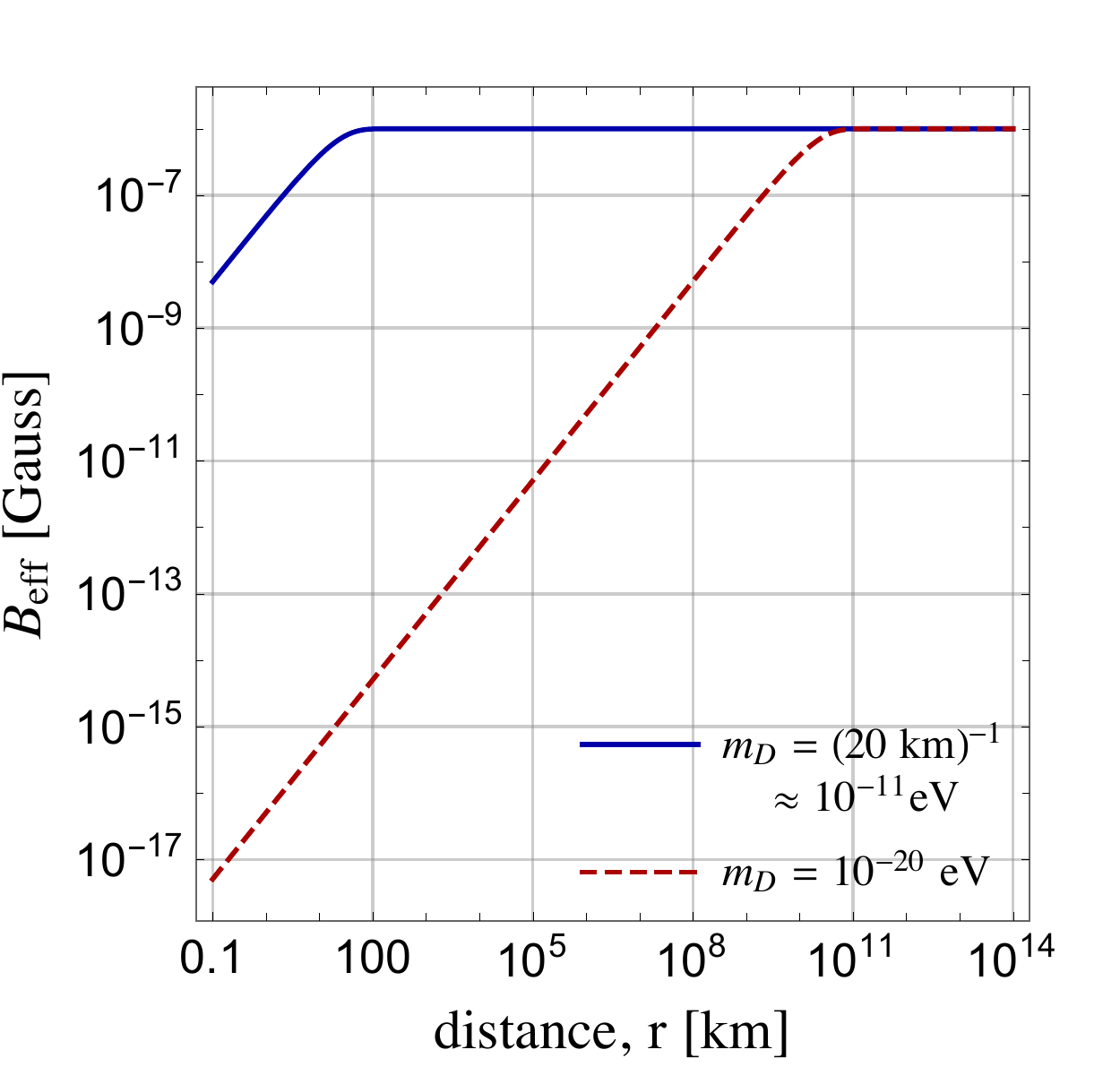}
   \caption{Radial profile of the galactic magnetic field felt by a {dark} monopole with $\varepsilon = 1$, for two choices of dark photon mass, {and constant galactic field $B=3 \times 10^{-6}$ G}. {Here $r$ is the distance between 
   the dark monopole and the source of an ordinary magnetic field.}}
   \label{fig:B}
\end{figure}

However, when the dark sector contains magnetic charges (i.e. magnetic monopoles), the opposite millicharging occurs, 
for at large distances dark magnetic charges 
acquire a millimagnetic charge under electromagnetism \cite{Hook:2017vyc} (see also these earlier papers which studied similar ideas~\cite{Brummer:2009cs,Bruemmer:2009ky,Sanchez:2011mf} as well as \cite{Terning:2018lsv,Terning:2018udc,Terning:2019bhg}). This feature is a consequence of the Dirac charge quantization condition which still holds for electric and magnetic charges at all scales, but only after summing over the two sectors \cite{Terning:2018udc}. Thus at large distances, magnetic currents of the dark sector 
will feel $\varepsilon$-suppressed magnetic fields \cite{Hook:2017vyc}
\be 
B_{{\rm eff}} = \varepsilon B \left(1-e^{-m_{D}r}     \right)~.
\label{eq:Beff}
\ee
Here $m_{D}$ is the dark gauge boson mass, $r$ is the distance from the dark magnetic charge to a source of 
a static magnetic field producing an electromagnetic field $B$ at $r$.  
At distances less than $m_D^{-1}$ to such a source, dark monopoles experience a magnetic field additionally 
suppressed by $m_{D}r$, but 
at larger distances they feel only a suppression by $\varepsilon$. As an example, we display the distance dependence on this in Fig.~\ref{fig:B}.

In this theory a Dirac quantization condition stills holds, but one must 
sum over magnetically and electrically charged pairs across both the dark and ordinary sectors. This means 
that it can violated between any two pairs. In particular, 
at large 
separations between a millimagnetic monopole and an ordinary electron, 
the Dirac quantization condition between these two particles is violated, since the magnetic flux tube connecting the monopole/anti-monopole pair 
acts as a physical Dirac string {of thickness ${\cal O}(m_D^{-1})$}. At shorter distances
the dark photon makes a contribution 
to the Aharonov-Bohm phase, making the phase unobservable \cite{Terning:2018udc}. 
The Aharanov-Bohm phase shifts induced by these flux tubes has been proposed as a viable method for 
testing millimagnetic dark matter, provided the experiment is large enough to completely enclose 
the magnetic flux tube~\cite{Terning:2019bhg}.


In this paper we examine the interaction between the galactic magnetic field and milli-magnetic charged particles (mmCPs), which we will assume comprises all of the DM. As we will see, in vacuum mmCPs can form either stable hydrogen-like or stringy-like bound states consisting of a confined monopole/anti-monopole pair, 
depending on the values of underlying magnetic coupling and dark photon and dark monopole masses.
The focus of this paper will be almost entirely on parameter values leading to atomic-like states, 
though we briefly comment on how some of our results change when the bound monopole system is stringy-like.

{Whether the bound monopole states are presently in their ground state is a separate question. As we shall see in 
Section \ref{sec:Relaxation-to-the-Ground-State}, for much of the parameter space of interest here, extended classical states 
and excited atomic states are extremely long lived. Moreover, the typical kinetic energies of the bound monopoles are 
much larger than their binding energies. Hence inelastic collisions between bound monopoles can produce 
long-lived excited states, dissipating the initial kinetic energy. A galactic population of dark monopoles 
is probably characterized by a collisionless, non-degenerate plasma, comprised of highly excited states and extended string states, rather than a collection of bound monopoles all in their ground state. Characterizing 
the different occupation numbers requires following the coupled Boltzmann equations, including dissipative processes, over the history of the Milky Way galaxy.}

This is a difficult problem and beyond the scope of this paper. Instead, we will assume that all of bound 
monopole systems are in their ground state, at least when it is self-consistent to do so. This will 
lead to a conservative upper bound on the monopole coupling, arising from bounds on self-interacting dark matter. The above assumption is not self-consistent  
when the interparticle spacing becomes smaller than the Bohr radius, and here we resort to approximating 
the galactic population as a non-degenerate, collisionless plasma.

In the presence of an external magnetic field, the atomic-like state is no longer the configuration of lowest energy. In analogy with the Stark effect as we explicitly show, the bound monopole/anti-monopole pair can quantum mechanically tunnel to a configuration on the other side of the 
Coulomb barrier, after which the constituents are accelerated apart by the background 
magnetic field. In the case of a plasma 
there is no barrier preventing the instability from forming, and so 
in the presence of an external magnetic field the monopole-anti-monopole pair are accelerated in opposite directions.

The energy added to the monopole/anti-monopole pair is compensated by a loss of galactic magnetic field energy. If this energy loss is sufficiently rapid, it could destroy the observed galactic magnetic fields, in contradiction with observations. 
For monopoles charged under electromagnetism, this is the well-known Parker effect \cite{Parker:1970xv}, 
which was further generalized in \cite{Turner:1982ag, Adams:1993fj}. The Parker bound is extremely stringent,  limiting ordinary monopoles having GUT-scale masses to less than $O(10^{-9})$ of the local dark matter energy density. For even less massive ordinary monopoles this bound becomes significantly stronger.
As we will see, in a similar way, the existence of galactic magnetic fields on long timescales places leading constraints on the mmCP parameter space. 

To arrive at these results, we first bound the self-coupling $g_D$ of the dark magnetic 
monopoles that arise from Bullet Cluster and halo ellipticity constraints. This analysis 
is self-contained and provides new, independent bounds on this coupling.

These bounds are 
then used in two independent ways as input into
bounds on the strength of the effective magnetic coupling $Q_m \propto g_D \varepsilon$ of the dark 
monopoles to ordinary magnetic fields. In the first, we simply use bounds on $\varepsilon$ obtained 
from other astrophysical and cosmological sources. While this leads to a strong 
constraint on $Q_m$, we find this bound is complementary, as for smaller monopole masses it is weaker than the bound obtained here from the Parker effect, though stronger at larger monopole masses.

We begin by discussing relevant properties of the bound monopoles. We discuss the conditions under which the bound monopoles are atomic-like. Next, we discuss bounds on the self-scattering of mmCPs. Then we continue with a discussion 
of the original Parker bound, modified for millimagnetic monopoles. We then discuss the quantum tunneling process for the ground state of the bound 
monopole/anti-monopole pair. Our main numerical results are shown in Figs.~\ref{fig:sidm} and ~\ref{fig:bounds}. We also summarize other 
relevant constraints, the most pertinent being from magnetars \cite{Hook:2017vyc}, which we find to be weaker.

\section{mmCP Properties}

Before we consider the milli-monopole system in the galactic magnetic field, we first discuss its properties in vacuum. 
We 
assume the dark photon is Higgsed with a mass $m_D$.

 We model a dark monopole and anti-monopole pair separated by a distance $r$ via the Hamiltonian
\be 
H_0 = \frac{p^{2}}{2\mu} - \frac{\alpha_{D}}{r}e^{-m_{D}r} + C \pi v_D^2 r~~ 
\ee
where $\mu=M/2$ is the reduced mass and $M$ is the mass of the monopole. 
$v_D$ is the vacuum expectation value (vev) of a dark Higgs field giving rise to a
confining force $C \pi v^2_D$ between the monopole pair, and $C$ is a dimensionless parameter of order one. 
The coupling $\alpha_D=g^2_D/4\pi=4 \pi/e^2_D$ is the dark magnetic fine structure constant, in terms of 
a fundamental dark electric charge $e_D$. (More details on our conventions, as well as the modified Maxwell's equations, are provided in Appendix \ref{app:notation}).
Here we assume the dark monopole and anti-monopole are of different flavors so that the bound monopole state is stable \cite{Terning:2018lsv}. 
For simplicity we further assume the two flavors have the same mass and have the same absolute minimal magnetic
charge in units of $4 \pi/e_D$. These assumptions do not qualitatively affect 
our conclusions in any way. This Hamiltonian captures the dynamics of the bound states provided the monopoles are non-relativistic, which occurs when $m_D\ll M$ and the magnetic Coulomb interactions are small.

The first potential term is the familiar magnetic Coulomb interaction. 
The second potential term dominates at large $r$ and models the
confinement of magnetic charge and the tension of the string 
connecting the monopole-anti-monopole pair \cite{Nielsen:1973cs,Nambu:1974zg}. {Each monopole/anti-monopole has a flux 
$\pm 2 \pi n/e_D$ and therefore can be supported at either end of the Nielsen-Olesen string that carries the same amount of flux.
The tension $T$ of the Nielsen-Olesen string is given by the vev $v_D$ and an $O(1)$ constant related to 
the amount of magnetic flux, $T\simeq v^2_D$. This result 
can be demonstrated for a specific value of the dark Higgs self-coupling $\lambda_D \propto e^2_D$, which in 
a supersymmetric version of the model would imply
the vortex is a BPS state 
\cite{Banks:2008tpa}. For 
more details we refer the reader to Appendix \ref{appendix:Nielsen-Olesen-vortex}. We will use this value 
for the string tension throughout.}

We will be most interested in the situation in which the ground and first few excited states are largely 
determined by the Coulomb interaction, 
rather than the confining force. This amounts to requiring the Bohr radius of the first 
few low-lying states be smaller than the de Broglie wavelength of the dark photon, or in other words, 
\be 
m_D \ll \alpha_D M ~.
\label{eq:atomic-like-condition}
\ee 
This condition is equivalent to using second-order perturbation theory to self-consistently treat the confining term as a small correction to the energies
and wavefunctions of the low-lying atomic-like states, {the latter of course being exponentially suppressed at 
large $r$}. 
Then the energy and length scale of the system follow the hydrogen-like results
\bea 
E_{0} &\simeq& -\frac{M}{4}\alpha_{D}^{2} \simeq - 3\times 10^{-23}~{\rm eV}~\left(\frac{M}{10^{4}~{\rm eV}}\right)~\left(\frac{\alpha_{D}}{10^{-13}}\right)^{2} \nonumber \\
L_{0} &\simeq & \frac{2}{M\alpha_{D}} \simeq 0.4~{\rm km}~\left(\frac{10^{4}~{\rm eV}}{M}\right)~\left(\frac{10^{-13}}{\alpha_{D}}\right), 
\label{eq:atomic}
\eea
where we have adopted a fiducial set of parameters $M=10^{4}$ eV and $\alpha_{D} =10^{-13}$. The choice of these fiducial values is in the middle of the range of interest. 

Conversely, for parameter values $\alpha_D M \ll  m_D \ll M$
the spectrum is determined by the confining potential and is more string-like. The wavefunctions are given by Airy functions, 
with typical size $L_{s} \sim \left(C \pi v^2_D M\right)^{-1/3}$ and  
energy eigenvalues  
$E_n \sim \left(C \pi v^2_D M (n+1/2)\right)^{2/3}/M$. This parametric dependence on the underlying 
parameters of the potential can be 
understood by rescaling the coordinates and energy to dimensionless units \cite{Terning:2019bhg}, or by 
imposing the Sommerfeld quantization condition \cite{Shankar:102017}.


\subsection{{Bohr radius, interparticle spacing, dark photon mass, and dark Debye length}}
\label{sec:length-scales}

A number of physically distinct length scales play a role in determining the relevant physics of the monopoles in a given situation. As discussed above, the monopole/anti-monopole bound states are characterized by the Bohr length in Eq.~\ref{eq:atomic}. In addition, we will assume that these bosonic states are ``particle-like'' which 
have low occupation numbers, rather than ``wave-like'' having large occupation numbers. This 
condition is true as long as the deBroglie length, $\lambda_{\rm dB} \sim (M v)^{-1}$, of the monopole is much smaller than the interparticle length, $\lambda_{\rm I} \sim n^{-1/3} \sim (\rho/2M)^{-1/3}$. Thus the monopoles will be particle-like in the Milky Way for masses, $M \gtrsim 10$ eV.  

\begin{figure*}[t!]
\includegraphics[angle=0,width=.5\textwidth]{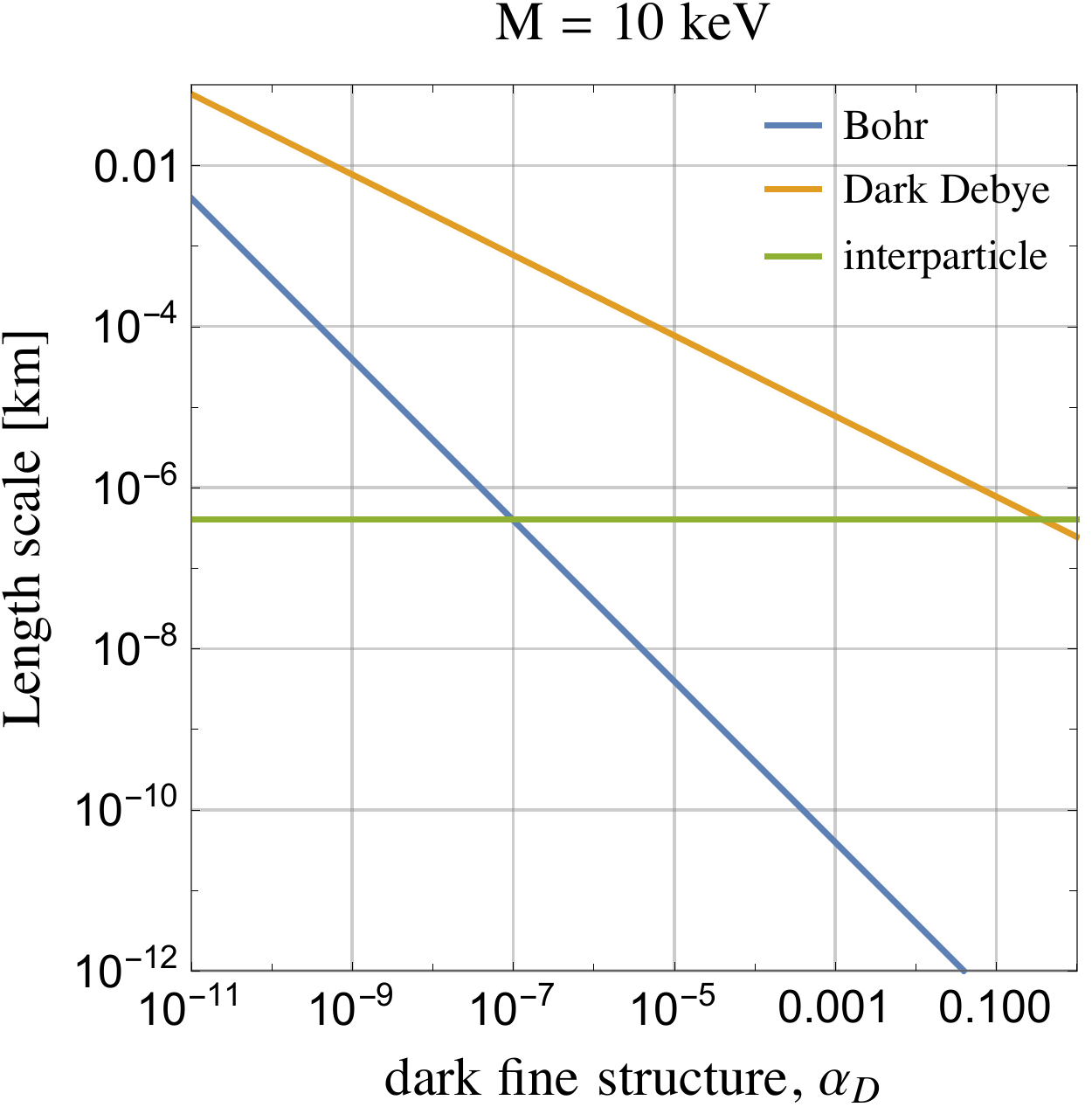}%

   \caption{Here we show the Bohr, {dark Debye}, and interparticle length scales for an example $M=10$ keV monopole as a function of the dark fine structure constant. {The notion of a 
   dark Debye length only exists for $\alpha_D \lesssim 10^{-7}$ as described in the text.}}
   \label{fig:lengths}
\end{figure*}

Next, if the density of monopoles is 
sufficiently large they do not form atomic-like states.
We expect this transition to occur when the interparticle length is smaller than the Bohr length. In Fig.~\ref{fig:lengths}, we show the interparticle and Bohr lengths as a function of the dark coupling for an example $M=10^4$ eV monopole. As a result, in this example one can see that for couplings $\alpha_{D} \gtrsim 10^{-7}$, monopoles in the galaxy are effectively atomic, whereas for $\alpha_{D} \lesssim 10^{-7}$, the monopoles 
are no longer bound into atom-like states.

At these smaller couplings, the collection of dark monopole and anti-monopole pairs instead form a classical, collisionless dark plasma. The plasma is non-degenerate, because as we saw above, we assume $M \gtrsim 10$ eV such 
that 
$n \lambda^3_{\rm dB} \ll 1$ holds. (Actually, throughout we will take 
$M \gtrsim 100$ eV.) The system is a plasma, with the magnetic charges screened 
at distances larger than the ``dark'' Debye length. We note that while the magnetic monopoles 
are confined, confinement of magnetic charge only occurs at distances larger 
than the thickness $m_D^{-1}$ of the magnetic flux tubes. This length scale will be assumed 
to be macroscopic throughout this paper, and is far larger than any other scale in the problem. In vacuum,  at 
distances less than $m^{-1}_D$ the dark magnetic charges are unscreened.

Now in this plasma, the typical Coulomb energy in a volume of size $\lambda_{\rm I}$ is assumed to be 
much smaller than the 
typical kinetic energy -- so the plasma is an almost ideal gas -- a condition that is equivalent to the dark Debye length being much larger 
than the interparticle spacing \cite{Landau:1980mil,Landau:Physical-Kinetics} (but it will 
still be less than $m_D^{-1}$).
Fig.~\ref{fig:lengths} displays the dark Debye length 
\be 
\lambda^{-2}_{\rm D} \simeq \frac{8 \pi \alpha_D }{\mu v^2}n_{\rm free} + m^2_D
\ee
associated with the magnetic Debye screening length within the dark plasma, for a reference 
monopole mass of $10$ keV. Note that for this reference mass, the notion of this physical scale makes no sense 
for $\alpha_D \gtrsim 10^{-7}$, for then the monopole/anti-monopole pairs are bound into atom-like states.
Here $n_{\rm free}$ is the number density of free monopoles or anti-monopoles, which is just $n$, since we make the simplifying assumption that once $\lambda_{\rm I} < L_0$ no monopole/anti-monopole pairs are in their ground state.
This length scale has played an important role in studies of dark matter self-interactions~\cite{CyrRacine:2012fz}. Over the range of couplings and masses that we will be interested in, this dark Debye length will not play a significant role. 

The important point here is that in the very weak coupling limit where the monopole/anti-monopole 
pairs are described by a classical, collisionless plasma, the presence of a background 
magnetic field (dark or ordinary), will cause the monopole pairs to be accelerated in opposite directions, provided, 
in the case of an ordinary magnetic field, 
the distance to the source is larger than $m^{-1}_D$. Unlike for the atomic-like states, here there is no 
quantum or classical barrier preventing acceleration.

Further, the dark photon implies an important length scale. Given that magnetic fields in the galaxy have a coherence length~\cite{Turner:1982ag}, $\ell = 0.3 $ kpc, we will assume that the dark photon is sufficiently heavy that the dark monopoles acquire a milli-magnetic charge under ordinary electromagnetism arising 
from Galactic sources (see Eq.~\ref{eq:Beff}). As such, we will require that the dark photon mass satisfies, $m_{D} \gtrsim \ell^{-1} \sim 10^{-26}$ eV. Additionally, we only 
require $m_D \ll \alpha_D M$ as described in the previous Section. For
our numerical results we will consider two reference values, $m_D = 10^{-20}$ eV 
and $m_D=10^{-10} \approx 10 ({\rm 20 km})^{-1}$ eV. The latter of these is chosen to correspond roughly with the size of a magnetar, which is a significant source of constraints on the model. The former value is a representative choice between $\ell^{-1}$ and $(20~{\rm km})^{-1}$. 
These choices only affect the magnetar bounds discussed in Section \ref{sec:magnetar-bounds}, 
and the independent bounds on $\varepsilon$ as discussed and quoted in Section \ref{sec:add}.

\subsection{Self-Interacting DM Bounds}

We can immediately determine some important constraints on the properties of this system by examining the bounds on the dark coupling arising from dark matter self-interactions~\cite{Spergel:1999mh,Tulin:2017ara}. 
The
momentum transfer or transport cross section is given by
\be 
\sigma_{T} = \int \frac{d \sigma}{d \cos \theta} (1-\cos \theta) d \cos \theta ~.
\ee

While a detailed examination of the self-interaction dynamics of the monopole/anti-monopole bound states is beyond the scope of this work, we can examine two limits of the self-scattering momentum-transfer cross section. First, given our focus on atom-like states we can use the results from Ref.~\cite{Cline:2013pca} which considered the 
elastic scattering between atomic DM states via an effective inter-atomic potential {that at long distances describes the van der Waals interaction
and at short distances describes Pauli repulsion.} They have obtained accurate fits to the full numerical solution which in {physical} 
units are of the form
\be 
\sigma_{T} = \pi L_{0}^{2} \left(b_{1} + b_{2} (E/E_{0}) + b_{3} (E/E_{0})^{2} \right)^{-1},
\label{eq:H-H-scattering}
\ee
where $E$ is the CM energy and $(b_{1}, b_{2}, b_{3} ) = (0.011,0.221,0.063)$ are 
numerical constants from the fit reported in Table I of Ref.~\cite{Cline:2013pca}. Observe 
that at low energies $E \ll E_0$ the momentum-transfer cross-section $\sigma_T \sim 10^2 \pi L^2_0$, 
that is enhanced above the na\"ive geometric cross-section, and also 
in agreement with \cite{CyrRacine:2012fz}. But for high-energies $E \gg E_0$ this cross-section 
is much suppressed, for then $\sigma_T \sim \alpha^2_D/M^2 v^4$.
It is perhaps surprising that for $E \gg E_0$, $\sigma_T$ is 
parametrically the same as Rutherford scattering, because 
this limit is undoubtedly  
outside the range of validity of the numerical solution of Ref.~\cite{Cline:2013pca}.

We are quick to note, however, that this numerical fit should however only be used somewhat qualitatively for bound monopole scattering, since it incorporates some 
physical 
features that simply do not apply to the bound monopole system. Specifically, the dark monopoles in the 
dark monopole bound state are bosons, so at short-distances the inter-atomic potential has 
no Pauli repulsion. Moreover, for scattering of hydrogen-like atoms there are spin-singlet and spin-triplet channels, 
which are not present for bound monopole scattering because they are all bosons. 

Yet in the 
high velocity limit of interest to us, the eikonal approximation gives the scattering amplitude 
as an integral over the impact parameter, weighted by a phase that depends on the scattering potential appropriate 
to that scale. 
Thus at impact parameters $b \gg L_0$, 
we expect this numerical solution to correctly characterize, at least 
qualitatively, elastic bound-state--bound-state scattering, since at those distances the interactions between $H-H$ or between two bound state monopoles are both described by similar 
van der Waals potentials.

We note that models which are truly hydrogen-like (in the sense that $\alpha_{D}$
${\gtrsim v}$, with $v$ a typical virial velocity) have at low-energies $k L_0 \ll 1$ (or equivalently, $E \ll E_0$), where $k \sim Mv$, 
so they result in cross sections $\sim 10^{10}$ larger than what is currently allowed 
because for $v \ll \alpha_D$ 
the momentum-transfer 
cross-section is quite large, $\sigma_T \sim 10^2 \pi L_0^2$. 
Though as argued in Ref.~\cite{Terning:2019bhg}, this may be allowed if monopoles only comprise 10$\%$ of the DM, {for then the elastic scattering cross-section is essentially 
unconstrained}~\cite{Fan:2013yva}.

On the other hand, as alluded to above, the parameter space of interest for us will be at such small coupling that the CM energy is large compared to the binding energy. Thus we are in the limit $k L_0 \gg 1$, that is, $E \gg E_0$. This is a qualitatively 
different regime from that considered in Refs. \cite{CyrRacine:2012fz,Fan:2013yva,Agrawal:2016quu,Terning:2019bhg}, which are interested in much larger couplings $\alpha \sim O(10^{-3}-10^{-2})$.
As we shall see below, for sufficiently small couplings the monopoles can comprise 100$\%$ of the DM while remaining consistent with the bounds from self-scattering. This is because the momentum-transfer cross-section becomes suppressed for the scattering of ``fast'' but still non-relativistic bound states.

At large distances the interaction between two bound monopole states is given by the van der Waals potential  
\be 
V_{\rm vdW} \sim - \frac{\alpha_D}{L_0} \frac{L^6_0}{r^6}
\ee 
This form is only valid for $m_D^{-1} \gg r \gg L_0$, below which the bound monopoles experience 
Coulomb interactions. 
Due to retardation effects, at distances $r \gg L_0/\alpha_D$ the van der Waals potential decreases further as $V_{\rm vdW} \sim -L^6_0/r^7$  \cite{Holstein:2001}. Finally at distances $r \gg m_D^{-1}$ the inter-bound-state potential is 
exponentially suppressed. 

One can estimate 
the bound-state--bound-state scattering cross-section arising from hard scattering by 
solving for the impact parameter $b$ at which the kinetic 
energy of the bound states are comparable to their potential energy \cite{Ackerman:mha}, given 
at these distances by the van der Waals potential. This consideration
gives $b \sim (\alpha^8_D/v^2)^{1/7} (L_0/\alpha_D)  \ll L_0/\alpha_D$ for the $r^{-7}$ potential, and $ b \sim L_0 (\alpha_D/v)^{1/3} \ll L_0 $
for the $r^{-6}$ potential, neither of which is a self-consistent approximation. At these 
velocities, hard 
scattering therefore probes the interior of the bound states. 

In the limit $v \gg \alpha_D$ the impulse approximation is valid, and hard collisions between two bound monopoles probe scattering 
of free monopole particles, as we now show. 
Consider a frame in which one of the bound monopole states is at rest, and 
the other incident with velocity $v$. The collision occurs over a time $T \sim R/v$, 
with $R \sim L_0$. During this time, 
the change in momentum $\Delta p$ imparted by the binding (Coulomb) force
$F$ of the stationary bound state to one of its constituent monopoles is 
\be 
\frac{\Delta p}{p} \simeq \frac{F T}{p} \simeq \frac{\alpha_D}{v} \ll 1 ~,
\ee
with $p \sim \alpha_D \mu$ a typical bound state momentum. The impulse of the binding force on the constituents 
of the moving bound state are similarly tiny. 
During the time of the collision the momenta of the scattering particles have therefore been 
little affected by their binding force, and so in this time the particles may be treated as being 
essentially free. In solving the Schr\"odinger equation this approximation amounts to ``freezing'' the 
momentum distribution of the bound state constituents during the collision process \cite{Newton:1982qc}. 
The bound state constituents will then undergo Rutherford scattering, and at high enough velocities 
the main effect of 
the bound state on elastic scattering will be to regulate the large diffractive peak in the forward direction due to the collinear 
divergence. Compared to hard scattering, this diffractive peak makes a parametrically 
similar contribution to the momentum-transfer cross-section. Below 
we reconsider more carefully the effect of soft scattering. 

To determine the bound-state--bound-state scattering cross-section entails solving the 
Schr\"odinger equation for a 4-body problem, 
which is beyond the scope of this paper. Instead, we will resort once more to a simple 
estimate, by
solving for the classical turning point as before, now here applied to the constituents with Coulomb interactions. 
This gives 
\be 
b \sim  \frac{g^2_D}{Mv^2} 
\ee
or in other words, 
\be 
b \sim \left(\frac{\alpha_D}{v}\right)^2 L_0 \ll L_0
\ee
which is now self-consistent. 
The momentum-transfer cross-section is then 
\be 
\sigma_T \sim \pi b^2 \sim \frac{8\pi \alpha_{D}^{2}}{M^{2} v^{4}}
\ee
This result is not surprising:
the constituent monopoles {are} probed directly in the scattering, and thus {the cross-section} approaches the Rutherford scattering limit.
%
Including the effect of the collinear singularity in the forward direction corrects the above formula 
by a logarithm:
\be
\sigma_{T} \sim \frac{8\pi \alpha_{D}^{2}}{M^{2} v^{4}}~\ln{\left[\csc^{2}(\theta_{{\rm min}} /2)\right]}
\label{eq:pp}
\ee
where the minimum scattering angle is
\be \csc^{2}(\theta_{{\rm min}} /2)  = \left(\frac{L \mu v^{2}}{\alpha_{D}} \right)^{2} + 1,
\ee
in terms of the length scale $L$ giving the maximum impact parameter.

\begin{figure*}[t!]
\includegraphics[angle=0,width=.32\textwidth]{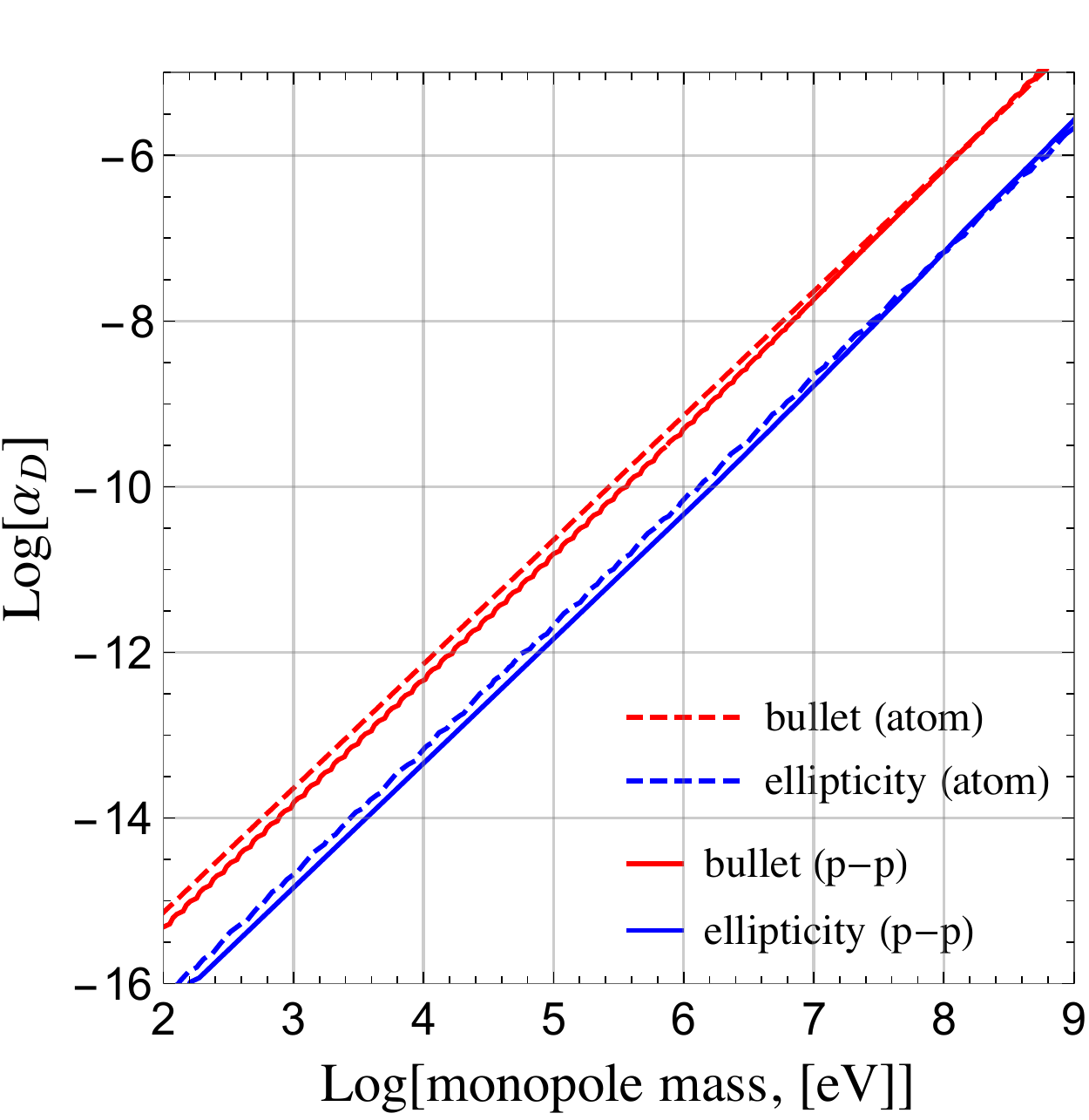}%
\includegraphics[angle=0,width=.34\textwidth]{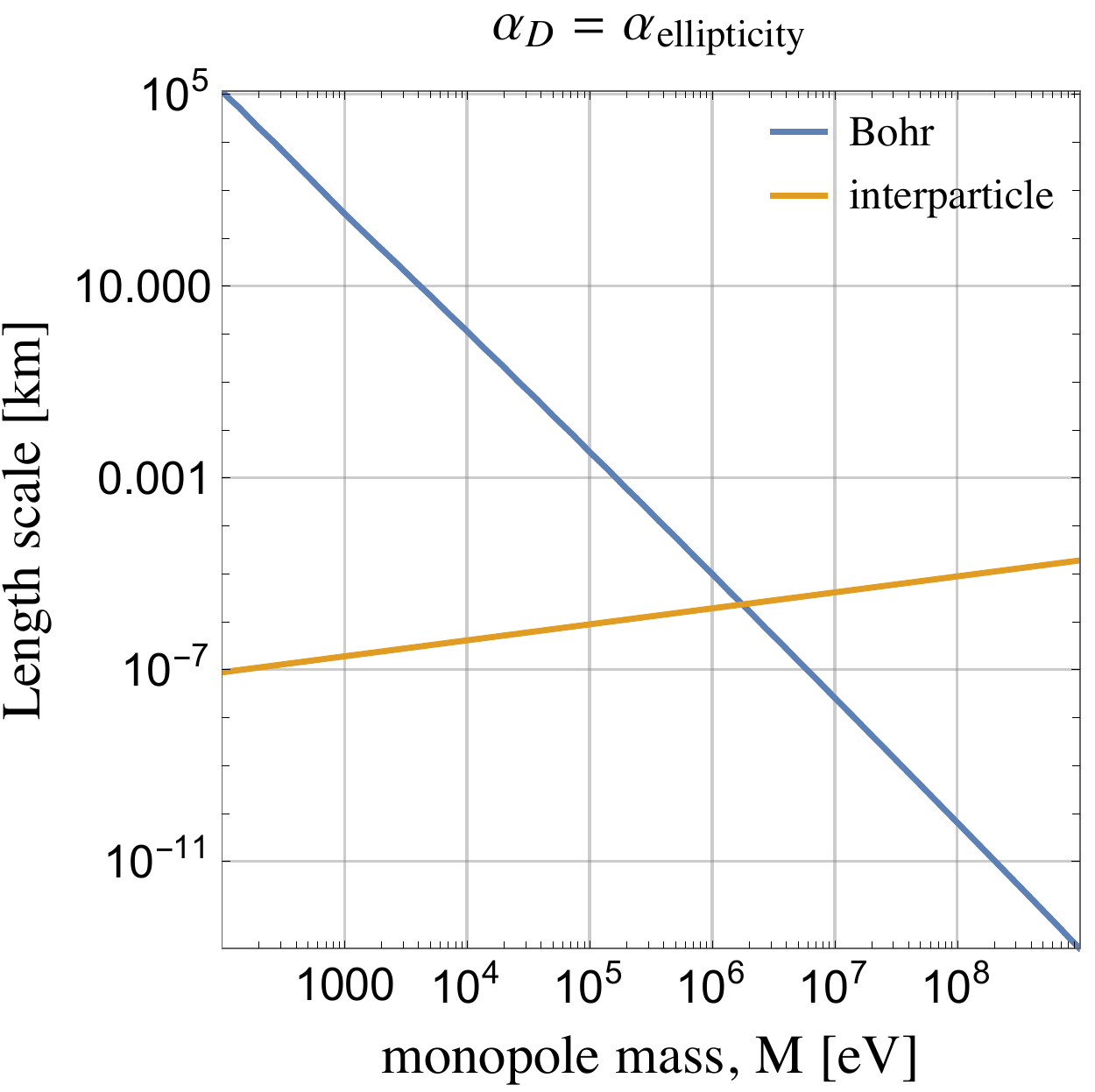}
\includegraphics[angle=0,width=.32\textwidth]{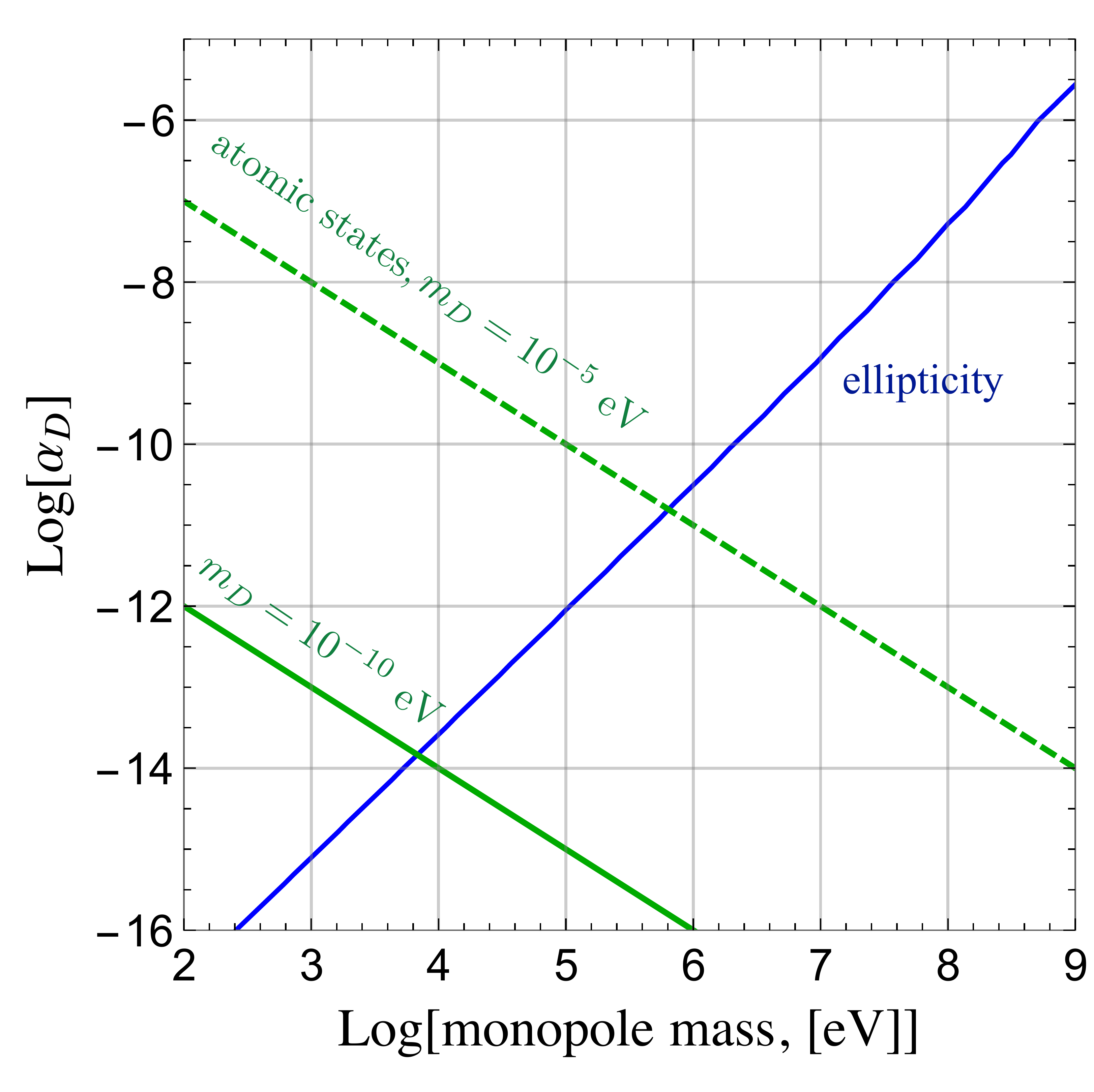}%
   \caption{In the {\it left panel} we show the resulting ellipticity  and bullet cluster bounds on the dark fine structure constant from DM self-scattering in the cases of atomic (Eq.~\ref{eq:H-H-scattering}) or $p-p$ scattering (Eq.~\ref{eq:pp}). In the {\it middle panel}, we show the Bohr and interparticle lengths at the dark coupling which saturates the strongest bound in the left panel. Lastly in the {\it right panel}, we compare the ellipticity bounds to the requirement that monopoles occupy atomic states (Eq.~\ref{eq:atomic-like-condition}) for two representative choices of dark photon mass.   }
   \label{fig:sidm}
\end{figure*}

For the case of bound monopole scattering discussed here, $L$ is simply the Bohr length. 
However, as discussed previously in Section \ref{sec:length-scales}, 
at small enough couplings the dark monopoles form a weakly coupled plasma instead of atoms, 
and in that circumstance $L$ is the smaller of the interparticle length and the dark Debye length, since 
Rutherford scattering is screened at those distances instead.
We thus consider three possible values of this effective length scale for scattering: the interparticle length, $\lambda_{\rm I}$, the dark Debye length, and the Bohr length. As an example, we display all three length scales as a function of the dark coupling in Fig.~\ref{fig:lengths}. Note that it is typically the Bohr length that is the smallest at large coupling and the inter-particle length which is smallest at small coupling. For the numerical results shown in Fig.~\ref{fig:sidm} and discussed below, we will always choose $L$ to be
the smallest of these scales. 

We now return to the issue of soft scattering, since in the discussion above for bound-monople-bound-monopole scattering, we treated the diffractive 
peak a little glibly. Each soft scatter contributes a small momentum-transfer $q$, but 
over many scatterings these can add up to a sizable change in kinetic energy. One way to estimate 
the contribution of the diffractive peak to the momentum-transfer cross-section is to approximately 
solve, in the fast limit, the classical equations of motion of a
a single monopole bound state moving in the van der Waals 
potential
of another bound state. This will give an estimate of the number of soft scatterings 
needed to cause a change in kinetic energy comparable to the initial kinetic 
energy. 
(This was done in Ref. \cite{Ackerman:mha} for the case of long-range 
Coulomb scattering.) 
We will see that 
soft scattering is parametrically the 
same size as the (hard) Rutherford scattering.

To first order in the transit time $T \simeq b/v$, the momentum transferred $\delta \vec{q}$ is approximately given by 
\be 
\delta q \simeq \pm V'_{\rm vdW}(b)  \left(\frac{b}{v}\right) 
\ee
with $\pm$ correlated with the azimuthal angle of the impact parameter $\vec{b}$, where recall 
$\vec{p}'_1=\vec{p}_1 + \delta \vec{q} $ and $\vec{p}'_2=\vec{p}_2 - \delta \vec{q}$ are the final momentum 
of the two scattered bound states, each having initial momentum $\vec{p}_1$ and $\vec{p}_2$. 
Following \cite{Ackerman:mha}, as each dark bound monopole state orbits the galaxy, it sees, roughly, a surface density of 
$\delta n = N/(\pi R^2) 2 \pi b db$ scatters, where $N$ is the number of dark matter particles and $R$ the size of the galaxy. 
Soft scatterings 
cause the DM particle to undergo a random walk, hence $\delta \vec{v}$, or equivalently, $\delta \vec{q}$ averages to zero, but $\langle \delta v^2 \rangle = (\delta v)^2 \delta n$ is non-vanishing. The impact parameter ranges 
from $b \sim L_0$ out to $\sim m^{-1}_D$, with by far the largest support occurring around $b \sim L_0$. It is 
therefore appropriate to use the $V_{\rm vdW} \sim r^{-6}$ potential, giving $\Delta v^2 \simeq (N/\pi R^2) (12 \alpha_D/Mv)^2$
for one orbit of the galaxy.
From this we infer the number of orbits $k$ needed to induce $\Delta v^2 \sim v^2$. Since a typical 
crossing time is $\tau_{\rm dyn} \sim 2 \pi R/v$, the time $\tau_{\rm soft}$ for these $k$ orbits to occur is then 
$k \tau_{\rm dyn}$. Empirically we require $\tau_{\rm soft}$ to be on the age of the galaxy (more on this below). Interpreting 
$\tau_{\rm soft} \simeq (\langle n \sigma_{\rm soft} v \rangle )^{-1}$ in terms of a soft-scattering cross-section 
$\sigma_{\rm soft}$, one finds 
\be 
\sigma_{\rm soft} \sim \frac{\alpha^2_D}{M^2 v^4} 
\ee 
as claimed, provided $v \gg \alpha_D$ which we assume throughout.


Next is the question of which physical systems set the most stringent bounds on DM self-interactions in our scenario. The two strongest constraints on our scenario come from the Bullet Cluster~\cite{Randall:2007ph} and halo ellipticity~\cite{MiraldaEscude:2000qt,Feng:2009hw,Peter:2012jh,Rocha:2012jg}. 
{There is another interesting constraint from requiring that dwarf galaxies not evaporate dark matter \cite{Kahlhoefer:2013dca}, which is stronger than the SIDM bound from the Bullet Cluster, but slight weaker than halo ellipticity. The 
 authors of \cite{Agrawal:2016quu} argue this bound is over-stated, and we will not include it here, 
 lying as it does between the other two bounds.} The halo ellipticity bounds arise from the effect of the DM self-interaction to isotropize the halo through the exchange of momentum. Consistent with previous work (e.g.~\cite{Agrawal:2016quu}) we find that halo ellipticity provides an order of magnitude stronger bound on the dark coupling than the Bullet Cluster. This is partly due to the fact our scenario predicts large cross sections at low velocities, and galactic halos exhibit velocities, $v_{gal} \sim 250~{\rm km}/{\rm s}$, which are substantially less than clusters, $v_{cl} \sim 1000~{\rm km}/{\rm s}$. We note that in our comparison we have assumed the bound given in the original Bullet Cluster paper, $\sigma_{T} < 0.7 ~$g$/$cm$^{2}$, ~\cite{Randall:2007ph}, although Ref.~\cite{Robertson:2016xjh} has argued that uncertainties in the unknown initial conditions of the colliding clusters may substantially weaken the bound.

To obtain the ellipticity bounds we compute the monopole-monopole scattering following~\cite{CyrRacine:2012fz} and requiring that the DM collision rate not allow more than 10 DM-DM scatters per Hubble time, $\Gamma_{{\rm col}} < 10 H_{0}$. We compute the collision rate by {numerically} integrating over the velocity distribution of DM, which we assume to be a Maxwellian with an average virial speed $v_{0} = 250~{\rm km}/{\rm s}$. 

In the left panel of Fig.~\ref{fig:sidm} we examine the impact of the atom-like or Rutherford-like (``proton-proton'') scattering assumptions on the Bullet cluster and halo ellipticity constraints. There we see that the constraints obtained from halo ellipticity in the Rutherford limit produce the most stringent bounds. In what follows we will use these bounds on the dark coupling, $\alpha_{D}$.

In the middle panel of Fig.~\ref{fig:sidm}, we plot the Bohr and interparticle length scales assuming that the coupling is at the upper bound set by the ellipticity constraints from the left panel. As we can see, the monopoles form a plasma for masses $\lesssim 2 \times 10^{4}$ eV, and are atomic-like for heavier masses. Note that this transition plays an important role when we consider the monopole evolution in a magnetic field since in the non-atomic limit they can be directly accelerated, but have to tunnel in the atomic case. We will return to the topic of tunneling in Sec.~\ref{sec:tunneling}.

In the right panel of Fig.~\ref{fig:sidm} we plot the ellipticity constraint along with the condition that the Coulomb interaction dominates over the confining potential, Eq.~\ref{eq:atomic-like-condition}. We have plotted the latter requirement for two choices of dark photon mass. Note that the coupling should exceed the value given on the dashed green curves in order to be in the Coulomb dominating regime.

Lastly, let us mention a few caveats. First, although we are mostly interested in ultra-light dark photons, we note that in some regions of parameter space these bounds may become somewhat weaker when $m_{D}^{-1}$ is smaller than the interparticle length. Second, as $m_D$ is further increased, 
eventually $\alpha_{D} M \ll m_{D}$ and the ground-states are not atom-like but stringy-like. In this limit the
self-interaction bounds are likely to be substantially modified. For then the typical size of the stringy ground state satisfies $L_s \gg m^{-1}_D$, so Coulomb interactions, and hence the van der Waals potential, are all exponentially suppressed at inter-particle distances. 
Any relevant scattering arises from impact parameters $b \sim L_s$. 
Na\"ively, the 
scattering cross-section is then simply $\sim L_s^2$. But 
to be more quantitative one
would have to include the dynamics of the Dirac string. This is because on scales $r$ 
with 
$ \sqrt{\alpha_D/(C \pi v_D^2)} e^{-m_D r/2} \ll r \lesssim  L_s$, the tension force between two monopoles
dominates over the Coulomb force, and this suggests that in considering the scattering of two monopole bound states one 
would have to better understand the role the flux tube plays.
And at smaller distances $r \ll \sqrt{\alpha_D/(C \pi v_D^2)} e^{-m_D r/2} $ Coulomb forces again dominate.

{More generally, for the effective Hamiltonian used here monopoles of different bound systems 
only interact through 
Coulomb forces and their derived van der Waals forces. This is because here the confining force has been treated as a rigid potential, acting only between monopoles of the same bound system. However, in a small impact parameter collision between two bound monopole 
systems, a monopole of one bound system could be ``swapped'' with a monopole of the other bound system. 
In studies of electron scattering off of hydrogen,  swapping become suppressed at energies high 
compared to the binding energy \cite{Newton:1982qc}, because the overlap of the initial and final state 
electron wavefunctions become suppressed. For monopoles to ``swap,'' the flux tube connecting them would 
have to break and reform. This is analogous to conventional magnetic reconnection, which to occur requires the presence of a 
current at the location of the break. Here, the only sources are the monopoles themselves, so a swap  
requires all four monopoles to be located at the same position. This effect is thus suppressed by phase space. 
The size of such a four-body contact interaction could be estimated 
by studying the low-energy effect of the non-linearities in the Higgs self-coupling and gauge interactions of the Nielsen-Olesen model.}

{We have also ignored inelastic processes, wherein 
\be 
MM + MM' \rightarrow MM^* + {MM^*}' ~.
\ee
Here say, two bound monopole states $MM$ and $MM'$ in their respective ground states, collide to form confined but excited states $MM^*$ and ${MM^*}'$. 
These processes are expected to not be suppressed compared to elastic scattering, for we are always in the limit 
$\alpha_D \ll v$
where the typical kinetic energies are much larger than the binding energy. This is a dissipative process 
and its effect on a population of dark monopole all in their ground state will not be explored here. Given 
that the excited states are long lived (see Section \ref{sec:Relaxation-to-the-Ground-State}) 
for the range of parameters of interest here, such processes will produce a meta-stable hetergeneous population of 
dark monopole states.}

Finally, we note 
that while we have focused on the momentum-transfer cross-section, that observable in itself may be 
too coarse \cite{Kahlhoefer:2013dca}. For this quantity
is insufficient 
to distinguish collisions that are soft but frequent and cause deceleration, as predicted in light mediator models, from rare but large 
momentum-transfer collisions caused by contact interactions. Further work in this area is clearly needed.

\subsection{Relaxation to the Ground State}
\label{sec:Relaxation-to-the-Ground-State}


The next important question is: how quickly is the ground state reached? The answer to this question is model-dependent, and relies on the detailed history of the production and evolution of dark monopoles in the early universe. 
For instance, the lifetime of a classical extended state
depends on the separation $L$ between the monopole and the anti-monopole, and the distribution for $L$ 
is contingent on the 
galactic and cosmological history. (The reader is referred to Ref. \cite{Sanchez:2011mf} for a discussion of one such distribution.)
As such, we will simply illustrate a few of the crucial timescales in sketching an answer to this question. 

We begin by considering a classical description of a monopole-antimonopole pair, in which 
they are widely separated by some distance $L$, with the magnetic 
flux tube connecting the pair approximated by a Nielsen-Olesen string \cite{Nielsen:1973cs}. However, $L$ must be smaller than a critical value (see Eq.~\ref{eq:pair} below), otherwise the free energy stored energy in the string can be lowered 
by producing a new monopole-anti-monopole pair. In vacuum, each monopole is accelerated by the tension in the string.  

In order to estimate the lifetime of such a configuration of the pair we follow~\cite{Berezinsky:1997kd,PhysRevD.83.123524} and use the relativistic Larmor formula in order to obtain the power radiated by the system into photons  
\be
P_{{\rm larmor}} = \frac{2}{3} g_{D}^{2} \varepsilon^{2} a^{2} = \frac{2}{3}\frac{g_{D}^{4}g^{2}Q_{m}^{2}}{ (4\pi)^{4}} \frac{m_{D}^{4}}{M^{2}}
\ee
where $a = v_{D}^{2}/M$ is the acceleration experienced by each monopole, and $v_{D} = g_{D} m_{D}/(4\pi)$. 
Following \cite{Hook:2017vyc} we define $Q_{m} \equiv \varepsilon g_{D}/g$, the milli-magnetic charge of the monopole measured in units of $g$, 
with $g=4 \pi/e$ a SM magnetic constant. 

Next note that if the pair is separated larger 
than $r_{\rm pair}=2M/v_D^2$, then the stretched string has enough energy to produce a monopole-anti-monopole pair. So it reasonable 
to only consider pairs separated no more than $r_{\rm pair}$, a length scale we will use to  
estimate the longest time for classical radiative decay.
Then for an initial separation given by $r_{\rm pair}$, the total energy that can be radiated away is
the sum of the initial kinetic energy in the monopoles and the stored energy in the stretched string. Near the 
threshold for pair production, the available energy
is dominated by the energy in the string, namely $2M$. We thus 
bound the lifetime in vacuum due to classical radiation into photons
as
\bea
\tau &<& P_{{\rm larmor}}^{-1} 2M \simeq \frac{3 (4\pi)^{2} M^{3}}{g^{2} Q_{m}^{2} \alpha_{D}^{2} m_{D}^{4}} \\
&\simeq& 10^{107}~{\rm yr}~\left(\frac{M}{10^4 {\rm eV}}\right)^3 \left(\frac{10^{-13}}{\alpha_D}\right)^2 ~\left(\frac{10^{-26}}{Q_{m}}\right)^2 ~\left(\frac{10^{-10}~{\rm eV}}{m_{D}}\right)^4,
\eea
which is many orders of magnitude larger than the age of the Universe for the parameters we consider. Of course, if the initial separation is smaller, then the lifetime may be much smaller.

In regions of parameter space with {$m_D v \gg a$}, radiation into dark photons will be exponentially suppressed 
~\cite{Berezinsky:1997kd}. However even in the limit that it is not, the Larmor power into dark radiation can be estimated as, $P^{(D)}_{{\rm Larmor}} \simeq 2 g_{D}^{2} a^{2}/3$, {with acceleration $a$ provided 
by the tension, as before.} This implies a lifetime limited by 
 \bea
 \tau_{D} &\lesssim & 12 \pi M^{3}/(\alpha^3_D m^4_D)  \\
  &\simeq& 10^{70} \left(\frac{M}{10^4 {\rm eV}}\right)^3 \left(\frac{10^{-13}}{\alpha_D}\right)^3 \left(\frac{10^{-10} {\rm eV}}{m_D}\right)^4 ~{\rm years} 
 \eea
We note again that this decay time can be much shorter if the initial monopole separation is small compared to $2M/v^2_D$.


Now that we have sketched out some of the classical evolution of the excited string, next we estimate the lifetime of the first excited state, assuming the dark photon mass is less than the typical difference 
in bound state energies. Assuming again that the monopole bound state is hydrogen-like, we estimate that the lifetime of the $2p \rightarrow 1s$ transition is~\cite{Shankar:102017} 
\bea
\tau_{2\rightarrow 1} &=& \left(\frac{3}{2}\right)^8\frac{1}{\alpha_{D}^{5} \mu} \\
&\simeq&  10^{40}~{\rm yr}~\left(\frac{10^{4}~{\rm eV}}{M}\right)~\left(\frac{10^{-13}}{\alpha_{D}}\right)^{5},
\eea
where in the second line we have assumed the value of the coupling for that fiducial mass that saturates the ellipticity bound from DM self-scattering. 

Thus if the monopoles are produced in some excited states in the early universe, they will remain in them by today, 
at least for parameter values around our fiducial values and below. 
However, to remain as general as possible regarding specific cosmological histories and production mechanisms, we will describe the bounds on the magnetic monopoles in the galactic magnetic field in a general framework which will allow us to understand how the bounds change depending on whether they are predominantly in the ground state or an excited state. 

Note that the dark photon itself will also be very long-lived. For sub-eV dark photons, the dominant decay channel is $\gamma_{D} \rightarrow 3 \gamma$. However given the strong astrophysical bounds on the kinetic mixing, this decay channel is vastly longer than the age of the universe for a $10^{-10}$ eV dark photon~\cite{McDermott:2017qcg}.

\section{mmCPs in Galactic Magnetic Fields: A Reformulated Parker Bound}


The presence of a galactic magnetic magnetic field in the Milky Way galaxy provides a constraint on the existence of millimagnetically charged particles. 
Basically, galactic magnetic fields accelerate monopoles, and the gained kinetic energy of the monopoles comes at the expense of a corresponding loss of energy in the magnetic field. For ordinary monopoles, a constraint on the number density of monopoles arises from 
requiring the galactic magnetic field not be dissipated faster than currents can regenerate it via 
dynamo action \cite{Parker:1970xv}.
For millimagnetic charged monopoles, the same considerations place an upper bound on its
effective magnetic coupling $Q_m$, assuming they comprise all of the dark matter.

In the remainder of this Section we bound the monopole flux by
paralleling the discussion in \cite{Turner:1982ag}, replacing the charge of the monopole 
$g$ with the millicharge $g Q_m$. 
We also model the ordered or smooth component of the Milky Way magnetic field as a $B = 3 \mu$G field, having a coherence length of $\ell$ = 0.3 kpc, and we take the timescale of the dynamo action to be $\tau_{\rm dyn}  =10^{8}$~yr~\cite{Turner:1982ag}.

First consider the velocity that a magnetically charged object would acquire via a background static magnetic field. The magnetic force on the monopole would be
\be 
F= (g Q_{m}) B ~.
\ee
The acceleration of a magnetically charged particle across a coherent region $\ell$ will impart to 
the monopole an energy
\be
\Delta E = g Q_{m} B \ell
\label{eq:energy_gain}
\ee
or equivalently a velocity
\begin{equation}
v_{{\rm mag}} \simeq \sqrt{\frac{2 \, g \, Q_{m} B\ell}{M}},
\label{eq:v_mag}
\end{equation}
assuming that the monopole was initially at rest. 
The relative size of $v_{mag}$ compared to the typical galactic velocities of virialized dark matter
is an important criteria for determining their evolution~\cite{Turner:1982ag}. 
As we will see, our region of interest is in the regime that $v_{{\rm mag}}$ is small compared to the typical galactic virial velocities $v \sim 200~ {\rm km}/{\rm s}$. Thus the velocities acquired by 
the monopoles due to the magnetic acceleration can be considered as small deflections to their virial velocity.

For a monopole in a uniform magnetic field, the equation of motion will be
\begin{equation}
M \frac{d\vec{v}}{dt}=g Q_{m}  \vec{B} ~.
\end{equation}  
This directly implies that the the kinetic energy $T\equiv (1/2) M v^{2}$ evolves via, 
$d T/dt = g Q_{m} (\vec{v} \cdot \vec{B}).$ However for two reasons we expect that the distribution-averaged change in kinetic energy, $\langle dT /dt \rangle$, vanishes. First the incident monopole distribution should be roughly isotropic. Second we expect equal fluxes of monopoles and anti-monopoles. Thus in order to determine the distribution-averaged change in energy we must go to second order
\be 
\langle T_{f} \rangle = \langle T_{i} \rangle + \frac{1}{2} (\Delta t)^{2} \frac{d^{2} T}{dt^{2}},
\ee
where $\Delta t  = \ell/v$. We can find a strictly positive expression for $d^2 T /dt^2$ by taking the time derivative of the equation of the motion, which for a static magnetic field yields
\be
M \frac{d^{2} T}{dt^{2}} = g^{2} Q_{m}^{2} B^{2},
\ee
or equivalently 
\be 
\langle \Delta E \rangle  = \frac{1}{2} (\Delta t)^{2} \frac{d^{2} T}{dt^{2}}= \frac{g^{2}Q_{m}^{2} B^{2}}{2M} \frac{l^{2}}{v^{2}} = \frac{1}{4} g Q_{m} B l \left(\frac{v_{mag}}{v}\right)^{2}.
\ee
%



Finally requiring that the monopoles and anti-monopoles 
not drain energy from the galactic magnetic field on a time scale shorter than the dynamo timescale $\tau_{dyn}$ implies,
\begin{equation}
\langle \Delta E \rangle  \times 2 {\cal  F} \times 4\pi \ell^{2} < \frac{B^{2}}{8 \pi} \frac{4 \pi}{3} \ell^{3} \frac{1}{\tau_{\rm dyn}},
\label{eq:Parker}
\end{equation}
in which the flux ${\cal F}$ corresponds to a flux of monopoles or anti-monopoles 
\begin{equation}
    {\cal F}= \frac{\rho_{M}}{2M} v,
\end{equation}
with the monopole mass density bounded from above by the local DM density, $\rho_{M} \le \rho_{DM} = 0.3~{\rm GeV}/{\rm cm} ^{3}$. These considerations imply the 
following upper bound on $Q_m$, 
\bea
Q_{m} &<& \sqrt{\frac{1}{12 \pi} \frac{1}{\tau_{\rm dyn}} \frac{v \, M^{2}}{ g^2 \rho_{DM}\, \ell} \left(\frac{\rho_{DM}}{\rho_M}\right)} \\
&\simeq& {4.5} \times 10^{-26}~\left(\frac{M}{10^{4}~{\rm eV}}\right)~\left(\frac{\rho_{DM}}{\rho_{M}}\right)^{1/2}
\label{eq:Parker_Q}.
\eea
Interestingly the magnetic field has cancelled out in the final result, though the coherence length $\ell$ and dynamo timescale $\tau_{{\rm dyn}}$ remain. The original work~\cite{Turner:1982ag} also observed that the Parker bound is independent of the magnetic field.

Note that while the bound obtained in Eq.~\ref{eq:Parker_Q} is quite strong at the low masses we consider, this is not inconsistent with the earlier bounds derived in Ref.~\cite{Turner:1982ag}. For the ordinary GUT scale monopoles with $Q_{m}=1$ and $M = 10^{16}$ GeV, the resulting bound on their density is of order $\lesssim 10^{-9}~\rho_{DM}$. Ultimately the Parker bound gets its strength from the extremely long distance, $\ell$, and time scales, $\tau_{{\rm dyn}}$, associated with the galactic magnetic field. 

{Lastly we note that magnetic plasma oscillations do not play a significant role in our calculations.  Ref.~\cite{Turner:1982ag} considered the possibility that the decay of the galactic magnetic field in the presence of monopoles, may only be the first half of an oscillation in the magnetic field. In this case, the monopoles themselves play a crucial role in the maintenance of the galactic magnetic field. However it was found in Ref.~\cite{1987ApJ...321..349P} that Landau damping is sufficiently efficient on kpc scales to obviate this potential revival of the decayed magnetic field.}
We refer the reader to Appendix~\ref{app:osc} for further details and numerical estimates.

%


\section{Tunneling from the Ground State}
\label{sec:tunneling}
Before the galactic magnetic field can start stretching the monopole pair apart, the system must first tunnel out of the ground state. To this end, we examine the Hamiltonian of the monopole pair in a background magnetic field. This problem is somewhat analogous to the Stark effect. Recall that we approximate the region of nonzero magnetic field in the galaxy as a box of uniform magnetic field.  

In the center of mass system, we use relative coordinates and model the system via the Hamiltonian 
\be 
H = H_0+g Q_{m} B(d) z,
\ee
where recall $B(d) = B \left( 1-e^{-m_{D}d} \right)$, and $d$ is the distance between the source and the bound monopole state. The magnetic field is static and oriented in the $z$-direction ($z=r \cos \theta)$. 
In this background the vacuum ground-state is only quasi-stable, and one can estimate the 
decay rate using one-dimensional instanton techniques \cite{Coleman:1985rnk}. While 
the problem is inherently 3-dimensional, in parabolic coordinates
\be
\xi=r+z~,~~ \eta=r-z ~
\ee
the instability is evident. For in these coordinates 
the potential is separable in the approximation that the Yukawa suppression is neglected. Then 
the instanton is clearly 
seen, since the one-dimensional potential for $\eta$ is unbounded from below at large $\eta$, provided 
the background magnetic field is larger than the tension. We refer the reader to Appendix \ref{appendix:parabolic-coordinates} for more details.

Here we note 
that the dependence of the decay rate on the magnetic field $B$ can be simply understood by recalling that the decay 
rate is \cite{Coleman:1985rnk}
\be 
\Gamma = K S_0 e^{- 2S_0} 
\ee 
where $S_0$ is the bounce action for the classical forbidden path, which in 
this case is $1/(3B)$ in natural units. The prefactor 
$K \propto \det (d^2/d \omega^2 + V^{''})$ depends on $\alpha_D$ but not on $B$, giving 
$\Gamma \propto B^{-1} \exp \left(-2 B^{-1}/3\right)$.

Under the WKB approximation one can also get the factors of $\alpha_D$ appearing in the prefactor $K$, and 
estimate the lifetime for the ground state to tunnel as~\cite{Landau:1991wop}
\bea 
\Gamma &=& \left( \frac{4 \alpha^5 _D \mu^3}{g Q_{m} B(d)}\right) e^{-\frac{2 \alpha^3_D \mu^2}{3 g Q_{m} B(d)}}, 
\label{eq:rate}
\eea 
where $\mu = M/2$.
Further details on the calculation of this decay rate are also provided in Appendix \ref{appendix:parabolic-coordinates}.

\begin{figure}[t!]
\includegraphics[angle=0,width=1\textwidth]{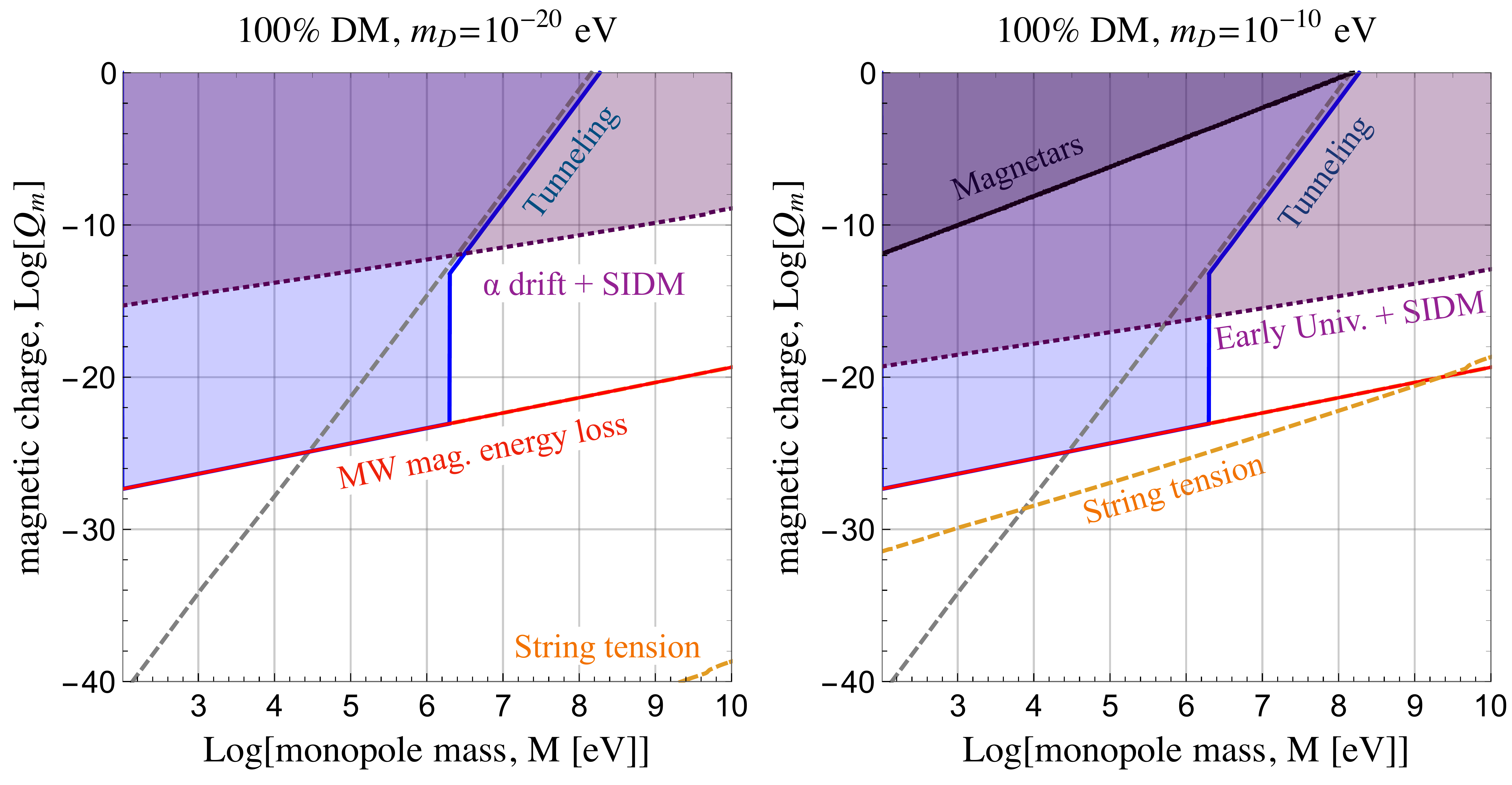}



   \caption{Here we display bounds on the magnetic charge $Q_{m}$ as a function of monopole mass for two choices of dark photon mass, $m_{D} = 10^{-20}$ eV (left) and $m_{D} = 10^{-10}$eV (right). Parameter space above the red curve can lead to excessive loss of energy of the MW's magnetic field. The region shaded in blue leads to tunneling out of the ground state on timescales less than the galactic magnetic dynamo, $\tau_{{\rm dyn}}= 10^{8}$ yr. In the region below the orange dashed curve, internal string tension dominates over the external magnetic force and no magnetic energy can be extracted from the galaxy. Thus the excluded region comes from the maximum of these three curves. 
   {If instead the population of monopole-anti-monopole pairs are described by a plasma, then in the background 
   Galactic magnetic field the monopoles are directly accelerated and the bound is instead given 
   by the red line.}
   For reference we also display the bounds from magnetar magnetic field lifetimes as derived in~\cite{Hook:2017vyc}. See text for discussion in Sec.~\ref{sec:add}
regarding the model-dependent bounds coming from the early universe and $\alpha$-drift. }
   \label{fig:bounds}
\end{figure}

The monopole parameters which lead to tunneling on timescales shorter than $\tau_{\rm dyn}$ are 
the regions lying about the blue curves shown in both panels of Fig.~\ref{fig:bounds}. 
To see that, first note that to obtain 
the upper bound for $Q_m$ for a given monopole mass $M$ shown in Fig.~\ref{fig:bounds}, we used as input the upper bound on $\alpha_D$ shown in the right panel of Fig.~\ref{fig:sidm},
obtained from requiring the bound monopoles pair satisfy halo ellipticity. Values for $\alpha_D$ weaker 
than this bound give a lifetime shorter than $\tau_{\rm dyn}$ -- because the potential barrier to tunneling is
shallower -- leading to a stronger bound on $Q_m$ than shown in the Figure. Values of $Q_m$ lying above the blue curve in Fig.~\ref{fig:bounds} are therefore excluded.

While we refer the reader to the Appendix \ref{appendix:parabolic-coordinates} for details of the calculation, we will summarize here a simple method for obtaining the approximate scaling of the tunneling rate. 
First, recall that within the WKB approximation the timescale for tunneling from the ground state 
can be written as
\be
\tau \propto \exp \left( 2 \int_{L_0}^{r_{\rm cl}}\sqrt{2 \mu \left(E_{0}-V(r)\right)} dr\right) \equiv e^{{2\gamma}},
\label{eq:rough}
\ee
where $L_{0} = 1/(\alpha_{D} \mu)$ and $E_{0} = \alpha_{D}^{2} \mu/2$, with $\mu=M/2$, are the ground state length scale and binding energy, respectively, and $r_{\rm cl}$ is the classical turning point, defined as the point at which the binding energy is equal to the external magnetic potential energy. 

Throughout our entire region of interest the classical turning point is approximately given by
\bea r_{{\rm cl}} &=& \frac{E_{0}}{g Q_{m} B(d)} = \frac{M \alpha_{D}^{2}}{4 g  Q_{m} B(d)} \\
&\simeq& 0.2~{\rm km}~\left(\frac{M}{10^{4}~{\rm eV}}\right)~\left(\frac{10^{-26}}{Q_{m}}\right)~\left(\frac{\alpha_{D}}{10^{-13}~{\rm eV}}\right)^{2}.
\eea
Moreover, given that integrand in Eq.~\ref{eq:rough} is well-approximated as linear in the vicinity of the turning point we find 
\be \gamma \simeq \frac{2}{3} \frac{p_{0}^{3}}{M F},
\ee
where $F= (g Q_{m}) B(d)$ is the external force (in our case the galactic magnetic field) and 
$p_{0} \simeq \alpha_{D} M/2$ is a typical momentum associated with the ground state. 

Use of the WKB semi-classical approximation is justified when the local de Broglie wavelength, $\lambda(x) \equiv 2\pi /p(x)$, does not vary appreciably, $|d \lambda(x)/dx| \ll 1$. This condition is equivalent to $m|F|/p^{3} \ll 1$, where 
$F$ is the external force (e.g., \cite{Landau:1991wop}). Notice that the validity of this approximation can equivalently be stated as requiring $\gamma \gg 1$. 

Given that the exponential term within $B(d)$ is approximately one in the region of interest, we find
\be \gamma \simeq \left(\frac{1}{12}\right)\frac{\alpha_{D}^{3}M^{2}}{ g Q_{m} B(d)}.
\label{eq:WKBvalid}
\ee
Using Eq.~\ref{eq:WKBvalid} we show the region of WKB validity ($\gamma \gg 1$) in Fig.~\ref{fig:bounds} as the dashed gray curve. Note as well that this agrees with the exponential factor in Eq.~\ref{eq:rate}. 

Finally we impose the condition that the decay rate (Eq.~\ref{eq:rate}) occurs on timescales less than the galactic dynamo, $\Gamma^{-1} \lesssim \tau_{{\rm dyn}}$. The exact solution is displayed in Fig.~\ref{fig:bounds} as the blue curve. However we can also arrive at an accurate analytic approximation iteratively by first noting that the exponential plays the leading role in setting the bound. As such, it is useful to define $Q_{m}^{*} \equiv \frac{\alpha_{D}^{3} M^{2}}{6 g B(d)}$, as the value of $Q_{m}$ for which the exponent in Eq.~\ref{eq:rate} becomes unity. 
We find that the bound is well approximated as 
%
\be
Q_{m} \lesssim \frac{-Q_{m}^{*}}{ {W}_{-1} \left(-\frac{1}{3\alpha_{D}^{2}M \tau_{{\rm dyn}}} \right)}, ~~~~~Q_{m}^{*} \equiv \frac{\alpha_{D}^{3} M^{2}}{6 g B(d)},
\ee
where $W_{-1}(x)$ is the $-1$ branch of the Lambert-$W$ function. Near $x = 0$, $W_{-1}(x)= \ln (-x) + O(\ln\ln (-x))$.

This approximation agrees quite well with the numerical implementation of the WKB integral shown in Fig.~\ref{fig:bounds}. Milli-magnetic monopoles with values of $Q_m$ in excess of this upper 
bound decay on timescales shorter than $\tau_{\rm dyn}$.

After tunneling, the monopole/anti-monopole pair is in an excited state (with respect to the vacuum), roughly 
separated by a distance given 
by the second classical turning radius $r_{\rm cl}$. {By definition of the second turning point, 
for $r> r_{\rm cl}$ the 
Coulomb force is weaker than the background magnetic force}, and 
the monopole and anti-monopole are subsequently accelerated 
in opposite directions, further increasing 
their relative separation. 

Next consider that once the monopole and anti-monopole pair are far enough apart, the energy arising from the string tension becomes sufficiently 
large so that pair production of monopoles will occur. This should occur roughly when $v_{D}^{2} r \sim 2 M$, or in other words at the length scale
\be 
r_{{\rm pair}} = \frac{2M}{v_{D}^{2}} = \frac{2M (4 \pi)^{2}}{g_{D}^{2}  m_{D}^{2}} \simeq 10^{29}~{\rm km}~\left(\frac{M}{10^{4}~{\rm eV}}\right)\left(\frac{10^{-13}}{\alpha_{D}}\right)\left(\frac{10^{-10}~{\rm eV}}{m_{D}}\right)^{2}
\label{eq:pair} 
\ee
Self-consistency of this analysis requires that we are always in the regime in which $r_{cl} < r_{pair}$, such that the classical turning point relevant for tunneling is small compared to the pair production length scale. This condition is equivalent to
\be
Q_{m} \gtrsim \frac{\alpha_{D}^{3} m_{D}^{2}}{32 \pi g_D B(d)} \simeq 10^{-48}~\left(\frac{\alpha_{D}}{10^{-13}}\right)^{5/2}\left(\frac{m_{D}}{10^{-10}~{\rm eV}}\right)^{2}~.
\ee
This condition is easily satisfied in the region of parameter space of interest, justifying {\it a posteriori} the tunneling computation.

Note that while the mechanism of continuous pair production in the galactic magnetic field could principle lead to the explosive growth of DM, we find that the time scale for this to occur is very long. This can be seen from the fact that the monopole pairs experience constant acceleration, $a_{mag} = g Q_{m} B(d)/M$ before reaching the distance $r_{\rm pair}$. Therefore the time-scale for pair production is $t_{\rm pair} = \sqrt{2 r_{\rm pair}/a_{mag}}$ or equivalently 
\be t_{\rm pair} = \frac{4 M}{m_{D}} \sqrt{\frac{\pi}{\alpha_{D} g Q_{m} B(d)}}.~~~~
\ee

For the light dark photon masses of interest here, this time scale is very large compared to the age of the Universe, but this growth in DM density in a background magnetic field could be relevant for larger dark photon masses. Also in our $m_{D} = 10 (20~{\rm km})^{-1}$  benchmark there are also regions of large $Q_{m}$ where $t_{\rm pair}$ can be short. However given that these regions are already excluded by the magnetic field energy loss arguments we do not consider this effect further in the present paper.

%



In Fig.~\ref{fig:bounds}, we display the regions 
where the tunneling decay time is shorter than $\tau_{\rm dyn}$ (above the blue curve), the regions in which the Milky Way's magnetic energy can be depleted (red curve), and the region below which the string tension dominates over the external magnetic field (orange dashed curve). This latter effect arises from the fact that for dark VEVs above a crtical value, $v_{D}^2 \gtrsim g \, Q_m\, B(d)$, the internal string tension prevents monopole acceleration. Since one needs all three effects to produce excessive magnetic energy depletion -- short tunneling time, sufficient magnetic energy loss, and weak enough string tension-- we display the newly excluded parameter space (shaded blue) as the region above the maximum of all three curves. Note that the magnetar bounds from~\cite{Hook:2017vyc} are also shown for reference in Fig.~\ref{fig:bounds}, and discussed in Sec.~\ref{sec:magnetar-bounds}. 

We also shown a purple dotted curve, which comes from combining the aforementioned SIDM bounds from halo ellipticity, $g_{D}^{(\rm max)}$, with the strongest bound on the kinetic mixing parameter $\varepsilon^{(max)}$ at each dark photon mass. One then 
immediately derives the upper bound on the magnetic charge
\be 
Q_m < g^{(\rm max)}_D \varepsilon^{\rm (max)}/g.
\ee
The kinetic mixing bounds are discussed in Sec.~\ref{sec:add}. From Fig.~\ref{fig:bounds} we see 
that at large mass this ``direct'' bound is stronger than the bound obtained from the Parker effect. The 
latter bound is weaker at larger monopole mass because of the cosmological 
assumption that all of the confined monopoles 
are in the ground state, which in the background Milky Way Galactic magnetic field has an increasingly longer 
lifetime to decay for increasing monopole masses.

Finally, notice that Fig.~\ref{fig:bounds} shows the maximum value for the string tension, which for a fixed monopole mass is consistent with halo ellipticity bounds. To understand 
that bound, note that we can express $T \sim m^2_D g^2_D$. Treating $g_D$ and $m_D$ as independent parameters, 
 then with $g_D$ bounded by halo ellipticity to lie below $g_{D}^{(\rm max)}$, $T$ 
 has its maximum value at $m^2_D (g_{D}^{(\rm max)})^{2}$. 
 Our analysis of the Parker effect is then self-consistent provided 
 $Q_m$ is larger than $m^2_D (g_{D}^{(\rm max)})^{2}/(g B(d))$.  Otherwise the internal string 
 tension is large enough to prevent both acceleration of the monopoles in a background magnetic field, 
 and quantum mechanical tunneling of the ground state. We note there is another 
 way to arrive at the same result, without having to treat $g_D$ and $m_D$ as independent parameters. 
 For 
an upper bound on $g_D$ is a lower bound on $e_D$. Then for a given fixed $m_D$, a lower bound on $e_D$ gives a 
maximum bound for $v_D$ or equivalently the tension $T$. In other words, for fixed $M$ and $m_D$, the upper bound on 
$T$ occurs when we saturate the upper bound on $g_D$.

\section{Magnetar and Other Independent Bounds}
\subsection{Magnetar Bounds from Schwinger Pair Production}
\label{sec:magnetar-bounds}


 In \cite{Hook:2017vyc} the authors consider the impact of milli-magnetic monopoles on magnetars. In particular, they compute the energy loss due to the magnetic equivalent of Schwinger pair production in the magnetar. As long as the internal monopole/antimonopole tension is sufficiently small, the pair will be accelerated apart. Note that the bounds derived from magnetars in this way apply generally, irrespective of whether or not the monopoles constitute DM.

Once the monopole-antimonopole pair is created, the system will extract energy from the magnetic field. This process will go on as long as the force coming from the string is smaller than the force pulling them apart
\begin{equation}
v_{D}^2 \lesssim g \, Q_m\, B(d).
\label{eq:stret_less}
\end{equation}

In this case, one derives a strong constraint on milli-monopoles simply from the fact that magnetar magnetic fields have existed for lifetimes at least $t_{{\rm life}} = 10^{4}$ years, over a region $d= 20$ km, with strength $B = 10^{15}$ Gauss.

This means, by conservation of energy
\begin{equation}
E_{\rm loss} \frac{g^2 Q_{m}^2\, B^{2}(d)}{ 4 \pi^{3}} \exp{\left[ \frac{-\pi M^{2}}{ g Q_{m} B(d)} \right]} < 
\frac{B^{2}}{2 \, t_{\rm life}},
\label{eq:magnetars}
\end{equation}
{where the energy loss is $E_{\rm loss} = g Q_{m} B(d) d$.}

We find that the magnetar bound for $m_{D} > (20~{\rm km})^{-1}$, is
\be 
    Q_{m}\lesssim 
\begin{cases}
    \left(\frac{2\pi^{3}}{t_{\rm life} d} \frac{1}{g^{3}B}\right)^{1/3}& M \lesssim  M^{*}\\
    \frac{\pi M^{2}}{g B},              & M \gtrsim  M^{*}
\end{cases}
\label{eq:magnetar-cases}
\ee
where $M^{*} \equiv \left(2 B^{2}/(t_{\rm life}d)\right)^{1/6}$. 
This bound is shown in Fig~\ref{fig:bounds} for a reference value  $m_D=10^{-10}~{\rm eV}$.  Also shown 
is the bound for a much smaller value of $m_D=10^{-20}~{\rm eV}$. Here the magnetar bound on $Q_m$ is somewhat weaker. 
The reason is that for this value, 
$m_{D} 
\ll d^{-1} \approx (20~{\rm km})^{-1}$, and consequently 
the effective magnetic field 
$B(d) \simeq B m_{D} d$ in (\ref{eq:magnetars}) becomes suppressed, and the Schwinger pair production 
rate is correspondingly much smaller. {To obtain these bounds we numerically solved 
Eq. \ref{eq:magnetars}, rather than using the approximate analytic solutions Eq. \ref{eq:magnetar-cases} which underestimates the bound on $Q_m$.}


\subsection{Additional Phenomenological Bounds On the Model}
\label{sec:add}

As discussed, the magnetar bounds are relevant for the parameter space of interest in this paper. However a variety of other constraints are relevant for other regions. 

\begin{itemize}

\item {\bf Cabrera SQUID detectors}

Superconducting loops have previously been used in the search for ordinary {bare} magnetic monopoles~\cite{PhysRevLett.48.1378}. The most stringent bounds from these detectors require that the flux of monopoles be $< 6.1 \times 10^{-10}~{\rm cm}^{-2}~{\rm s}^{-1}~{\rm sr}^{-1}$ for $Q_{m} > 0.06$~ \cite{PhysRevLett.48.1378}. Thus for monopoles comprising all of the DM, the superconducting loop bounds require large monopole masses, $M \gtrsim 10^{14}$ GeV as long as $Q_{m} > 0.06$. We note that precise bounds may have to be revisited since magnetic properties of bound monopoles are very different from ``free'' or ordinary monopoles.

\item {\bf MACRO}

The MACRO experiment performed a series of searches for GUT-scale monopoles using different detector sub-components~\cite{Ambrosio:2002qq}. Each of these searches was optimized for different incoming monopole speeds, $\beta$. Determining how these constraints map into our model of milli-magnetic DM is non-trivial. First we are interested in much lighter monopoles with vastly suppressed magnetic interactions. The stopping power of the mill-monopole pairs will be suppressed given that they are electrically and magnetically neutral, and the pair has no magnetic moment in the ground state.  In addition to the non-relativisitc dark matter flux of milli-monopoles, one may potentially also produce a flux of high-energy monopoles from cosmic ray scattering in the atmosphere. However this production may be more suppressed than naively expected since there is no $s-$channel production 
through dark and ordinary photon exchange \cite{Terning:2020dzg}.




\item {\bf Bounds on Dark Photon mass }

Even absent a coupling to monopoles, dark photons themselves produce an array of distinctive phenomenology (e.g.~\cite{Bauer:2018onh}). In our case however, we are mostly interested in ultralight dark photons in the range  $(10^{-10} - 10^{-20})$ eV. 
From reviewing the existing literature, we find that 
the most stringent bounds on the kinetic mixing arise from the early universe for $m_{D} = 10^{-10}$ eV and from an apparent ``drift'' in the fine-structure constant 
$\alpha$ for $m_{D} = 10^{-20}$ eV~\cite{Nelson:2011sf}. 
In Ref.~\cite{Nelson:2011sf} the authors consider the production of vector gauge bosons 
in the early universe from the misalignment mechanism. If these vectors thermalize with the SM via the kinetic mixing, they would acquire large energies compared to their masses and not constitute good cold dark matter candidates. 

The resulting large (oscillating) values of the background vector field induce time-varying electric and magnetic fields which are suppressed by the kinetic mixing parameter, $\varepsilon$.
The authors of Ref.~\cite{Nelson:2011sf} show that 
this can mimic the effect of a ``drift'' in the fine structure constant, which produces a 
bound at the $\varepsilon \sim 10^{-7}$ level for $m_{D} = 10^{-20}$ eV. 
This bound assumes that the dark photon contributes significantly to the background DM density, and is thus overly strong in our scenario. If the background vector density is 
vanishingly small however, the most stringent bound on kinetic mixing comes from the early universe which is constrained to be $\varepsilon \lesssim 10^{-4}$ at $m_{D} = 10^{-20}$ eV~\cite{Nelson:2011sf}.

\item {\bf Galactic magnetic seed and extension of the Parker Bound }

Monopoles drain energy from magnetic fields, inhibiting the growth of magnetic fields. Thus the growth of the galactic magnetic field from their ``seed'' fields may place stronger bounds than the ones explored here~\cite{Adams:1993fj}. Although this bound may be able to exceed the Parker bounds, it is subject to greater uncertainty. The origins of galactic magnetic fields remain quite controversial, 
with the basic underlying mechanism not well-understood. The values for the  
seed magnetic field and coherent length of the seed domain that give rise 
to the present-day galactic magnetic field span a very broad range. 
We refer the reader to \cite{2002RvMP...74..775W,2008RPPh...71d6901K} for recent reviews of this problem.

\item {\bf Aharonov-Bohm Direct Detection} 


In \cite{Terning:2019bhg}, it is suggested that milli-magnetic monopole DM can be probed via the phase shift they induce in Aharonov-Bohm type experiments. To induce the phase shift however, the monopole bound state must be in an excited state given that the ground state configuration has no dipole moment. Ref.~\cite{Terning:2019bhg} finds that solar photons can provide a mechanism to kick the monopole pair into an excited state if
\be  \frac{3}{16}\alpha_{D}^{2}  m_{D} \lesssim {\rm eV}
\ee

Next in order to remain in the excited state for a sufficiently long time, the energy difference between the excited and ground state must be small compared to $m_{D}$ so as to avoid prompt decay to the dark photon. This implies
\be m_{D} \gtrsim \frac{3}{16}M \alpha_{D}^{2}
\ee

Lastly, to avoid strong bounds on stellar energy loss from dark photon emission which decays to monopole pairs, Ref.~\cite{Terning:2019bhg} requires that the plasma induced photon mass be larger than twice the monopole mass ($m_{{\rm plasma}}\sim (1-20)~{\rm keV}$), such that this bound is completely absent. 

\item {\bf Symmetry Restoration}

Dark sectors with kinetic mixing to electromagnetism 
can have an even richer, potentially catastrophic astrophysical phenomenology. 
If
the dark Higgs boson is light, the $U(1)_D$ symmetry can be restored close to astrophysical sources having 
large magnetic fields \cite{Hook:2017vyc}.  Dark magnetic particles 
see $B_{\rm eff}$, which is suppressed close to the source. But as short distances 
$d \ll m^{-1}_D$, dark electric charged particles have an 
opposite physical behavior, for they see only an $\varepsilon-$suppressed magnetic field $\varepsilon B$. 
Thus for a source with compact support over distances $d \ll m^{-1}_D$, the effective potential for 
the dark Higgs field in the neighborhood of such a source has an effective mass 
\be 
m^2_{H,\rm eff} = m^2_{H,\rm vac}+|\varepsilon e_D B| 
\ee
which may be understood as arising from the quantization of a spin-0 particle 
in a background magnetic field (i.e., Landau levels) \cite{Ambjorn:1989sz,Ambjorn:1989bd,Ambjorn:1992ca,Heisenberg:1935qt,Maldacena:2020skw}.  
Inside spatial regions where $m^2_{H,\rm eff}$ is positive, the dark $U(1)_D$ symmetry is restored 
and the vacuum dark photon mass vanishes, though a plasma induced mass will still arise in analogy to the photon. It is difficult to speculate on the physical properties of such an exotic phase.
Such phases can be avoided if the physical mass of the dark Higgs boson is heavy enough, 
\bea
m_{H,\rm phys} &\gtrsim& \sqrt{\varepsilon e_D B} \nonumber \\
 &=& 3 \times 10^{-7} {\rm eV} \left(\frac{e_D\varepsilon}{10^{-6}} \frac{B}{3 \mu {\rm G}} \right)^{1/2}~,
\eea
a condition that is contingent on the source. 
As an example, in the atmosphere of a magnetar $B \sim 10^{15}$ G, the bound is 
$\simeq 10$ {\rm eV} $(e_D \varepsilon/10^{-6})^{1/2}$, and as mentioned above, only exists provided 
the inverse dark photon mass is larger than the size of the magnetar, or 
$m^{-1}_D \gg 20 {\rm km}$.

These brief comments 
are clearly independent of whether the spectrum has dark monopoles. 
\end{itemize}

\section{Conclusion}

We have reexamined the Parker bound on milli-charged monopoles. We have found that for a wide range of monopole and dark photon masses, the energy extraction from Galactic magnetic fields can place the leading constraints on the model. In particular, these constraints remain strong at dark monopole masses much greater than an eV, where the magnetar bounds become increasingly suppressed.

{To arrive at these bounds we have used as input halo ellipticity and Bullet Cluster bounds to constrain the self-coupling of the dark monopoles, 
finding the former consideration leads to roughly an order of magnitude stronger constraint.}

One might speculate on the cosmological origin of these particles. 
Monopoles can naturally arise in the early Universe through the Kibble-Zurek mechanism 
\cite{Kibble:1976sj,Kibble:1980mv,Zurek:1985qw,Zurek:1993ek,Zurek:1996sj}
during a second order phase transition 
in the early Universe. However, if the monopoles are to entirely comprise the dark 
matter, then their mass must be PeV or larger \cite{Murayama:2009nj,Graesser:2020hiv}, 
at least in the simple cosmological scenarios considered therein. These references also assume a 
na{\"i}ve relation between the monopole mass and critical temperature of the phase transition, motivated 
by classical Grand Unified Theories. In Seiberg-Witten theories and its generalizations \cite{Seiberg:1994rs,Seiberg:1994aj}, such a mass-temperature 
relation doesn't exist, and the monopoles can have masses arbitrarily smaller than the temperature of the phase
transition. This observation may hint at a necessary ingredient in any underlying particle physics 
construction for a Kibble-Zurek cosmological origin of such low mass monopoles. {And for 
such light dark photons considered here, bounds from the CMB are significant, essentially 
requiring the dark photons to never have been in thermal equilibrium with the SM in the early Universe.}

{For much of parameter space of interest here, the lifetimes for extended string 
states or excited atomic states to decay can be much longer than the age of the Galaxy. 
This raises the question of 
whether monopoles in the Galaxy ever cool to form atoms, or whether they always remain in the plasma state. 
Moreover, the typical kinetic energies of the bound monopoles greatly exceed their binding energies, 
and thus excited states can be re-populated through inelastic collisions. It would be interesting 
to explore, along the lines of \cite{Fan:2013yva},  whether these and other dissipative processes could cause a fraction of the dark monopole 
dark matter to form a galactic disk.}

Lastly, the direct detection of mmCPs is a topic deserving of further study. While Ref.~\cite{Terning:2018lsv} focused on using Aharonov-Bohm experiments to detect mmCPs, it would be worthwhile to examine the extent to which both MACRO~\cite{Ambrosio:2002qq} and conventional direct detection experiments could provide complementary sensitivity to mmCPs.

\section*{Acknowledgements}
We are very grateful for helpful discussions with Daniele S.M. Alves, Tom Banks, Jim Cline, Anson Hook, Patrick Huber, 
Hui Li, Duff Neill, and John Terning. 
The work of M.L.G is supported by the U.S. Department of Energy under the award number DE-AC52-06NA25396.
The work of I.M.S. and N.T.A. is supported by the U.S. Department of Energy under the award number DE- SC0020250. 

\newpage

\appendix
\section{Maxwell Equations with magnetic currents}
\label{app:notation}    
Following \cite{Coleman:1982cx} we choose our conventions of electric and magnetic charge so that
an electric charged particle with charge $q$ in units of $e$ and a magnetic charged 
particle with charge $g$ in units of $4 \pi/e$ give rise to electric and magnetic fields 
\bea 
\overrightarrow{E} &=& \frac{qe}{4 \pi} \frac{\hat{r}}{r^2} \longleftrightarrow J^0(x)= e q \delta^{3}(x) \\
\overrightarrow{B} &=& \frac{g}{e} \frac{\hat{r}}{r^2} \longleftrightarrow K^0(x)= \frac{4 \pi}{e} g \delta^{3}(x)
\eea
for the electric and magnetic charge densities $J^0(x)$ and $K^0(x)$.
Then before turning on kinetic mixing, the Dirac charge quantization condition is 
\be 
q g = \frac{n}{2}, ~n \in \mathbb{Z} ~.
\ee
The magnetic Coulomb potential between two monopoles of charge $g_1$, $g_2$, each in units of $4 \pi/e$ is then 
\bea 
V(r) &=& g_1 g_2 \left(\frac{4 \pi}{e^2}\right) \frac{1}{r} \\
 &=& g_1 g_2 \alpha_M \frac{1}{r}
\eea
where we have introduced the magnetic fine structure constant $\alpha_M=4 \pi/e^2$. 
This magnetic fine structure constant is none other than $\alpha_M=g^2/4\pi$, in terms of 
a fundamental magnetic charge $g=4 \pi/e$, and is not independent from $e$.

To understand how the milli-magnetically charged particles appear through the kinetic mixing we need to go to the Maxwell equations, which for our case are
\begin{align*}
    \partial_{\mu} F^{\mu \nu} - \varepsilon \partial_\mu F^{\mu \nu}_D &= eJ^{\nu}~, \\
        \partial_{\mu} F^{\mu \nu}_D  - \varepsilon \partial_\mu F^{\mu \nu} &=   e_D J^{\nu}_D + m_{D}^{2}   A^{\nu}_D ~,\\
        \partial_{\mu} \Tilde{F}^{\mu \nu} &=g K^\nu ~,\\
        \partial_{\mu} \Tilde{F}^{\mu \nu}_D  &= g_D K^{\nu}_D  ,
\end{align*}
where $\Tilde{F}_{\mu \nu}=\frac{1}{2} \epsilon_{\mu\nu\rho\sigma} F ^{\rho\sigma}$, 
$J$, $J_D$, $K$, $K_D$, are electric and magnetic currents for ordinary and dark sources.
Here $e$ and $e_D$ denote the electric charge for the ordinary and dark sector, and  
$g=4 \pi / e$, $g_D=4 \pi / e_{D}$ are the fundamental magnetic charges in each sector when $\varepsilon=0$.

To go to the mass basis one redefines the ordinary Maxwell gauge potential $A\xrightarrow{} A+ \varepsilon A_D$.
Then, Maxwell's equations at leading order in $\varepsilon$ are
\begin{align*}
        \partial_{\mu} F^{\mu \nu} &= e J^{\mu} ~,\\
        \partial_{\mu} F^{\mu \nu}_D  &=  e_{D} J^{\nu}_D+ m_{D}^2 \, A^{\nu}_D  + \varepsilon e J^{\nu} ~,\\
        \partial_{\mu} \Tilde{F}^{\mu \nu} &= g K^\nu- \varepsilon g_{D} K^{\nu}_D ~,\\
        \partial_{\mu} \Tilde{F}^{\mu \nu}_D  &= g_{D} K^{\nu}_D ~.
\end{align*}
In the physical mass basis, a dark magnetically charged particle has a direct coupling to a dark magnetic field 
$B_D$, but also an $\varepsilon$-suppressed coupling to an ordinary magentic field $B$. By the second equation above, 
an ordinary current creates a dark magnetic field $B_D$ that is $\varepsilon$-suppressed. But 
at large distances to the source this becomes exponentially suppressed leaving only the direct coupling to 
an ordinary magnetic field $B$ from the third equation. 
The effective magnetic field felt by a probe dark monopole sourced by an ordinary magnetic field is thus \cite{Hook:2017vyc}
\begin{equation}
    B_{\rm eff}(d)= \varepsilon B (1- e^{-m_{D} d})~.
\end{equation}
The dark monopole couples to the ordinary magnetic field with a milli-charge strength 
$\varepsilon g_{D} \equiv Q_{m} g$ (see \cite{Hook:2017vyc}, and also \cite{Terning:2018udc, Terning:2019bhg} which has a 
derivation of these results using the two-potential Lagrangian formalism of Zwanziger \cite{Zwanziger:1970hk,Schwarz:1993vs}).

{Finally, we comment on the Dirac charge quantization condition in the presence of kinetic mixing. 
In the physical mass basis the ordinary photon couples to $J$ and $K- \varepsilon K_D$, while the physical dark 
photon couples to $J_D + \varepsilon J$ and $K_D$. Thus the electric and magnetic 
charges associated to these currents cannot possibly satisfy the Dirac charge quantization condition 
separately in each sector. However, a `diagonal'  Dirac charge 
quantization condition is preserved \cite{Terning:2018udc} 
\be 
gq + g_D q_D= \frac{n}{2}, ~n \in \mathbb{Z} ~,
\ee
since evidently $\varepsilon J K_D$ cancels between the two terms summed. 
}

\section{Nielsen-Olesen vortex at the BPS point} 
\label{appendix:Nielsen-Olesen-vortex}
In this section we summarize some properties of the Nielsen-Olesen vortex/string solution \cite{Nielsen:1973cs}. 
We largely follow the cogent discussion 
in \cite{Banks:2008tpa}. 
The goal here is to show, at least for a special value of the self-coupling $\lambda$ of the Higgs boson, 
the tension $T$ of the solution is simply $T = 2 \pi n v^2$,

The energy density of the string is given by 
\be 
\mathcal{E}= \frac{1}{e^2} \left[ \frac{F^2_{12}}{2}+ |D_i \phi|^2 + \frac{\lambda}{2}(|\phi|^2 - \phi_0^2)^2 \right]
\ee
and the solutions are assumed to be independent of the $z$-direction. 
Note the uncanonical normalization of both the gauge field and the scalar $\phi$, so the only appearance of 
the electric coupling is as a universal factor. The vacuum is $\phi=\phi_0$ 
and the physical gauge boson mass is $m=\phi_0$. For the special value $\lambda=1$, 
the energy density can be rewritten as \cite{Banks:2008tpa}
\be 
\mathcal{E}= \frac{1}{e^2} \left[ \frac{1}{2}(F_{12}+ |\phi|^2 - \phi^2_0)^2 + \frac{1}{2}|D_i \phi+ \epsilon_{ij} D_j \phi|^2 +
\phi^2_0 F_{12} - i \epsilon_{ij} \partial_i (\phi^* D_j \phi) \right] 
\ee
The point is that, following Bogomolnyi-Prasad-Sommerfeld, here the Nielsen-Olesen string describing 
$n$ units of flux is found by simply minimizing the energy. First order equations for the string profile, which can be found in \cite{Banks:2008tpa}, are obtained by enforcing that each perfect square in the expression for $\mathcal{E}$ vanishes exactly. 
The last term is a total derivative. Since the radial profile 
of $\phi$ can be shown to approach $\phi_0$ at spatial infinity as $e^{-\sqrt{2} \phi_0 r}$, the only contribution to the 
energy density arises from the third term, 
\be 
\mathcal{E}= \frac{1}{e^2} F_{12} \phi^2_0 ~.
\ee 
The string tension is then 
\bea 
T &=& \int d^2 x \mathcal{E} \\
 &=& \frac{1}{e^2} \phi^2 _0 (2 \pi n) \\
 &=& 2 \pi n v^2
 \eea 
 since $\int d^2 x F_{12}=2\pi n$ is an exact expression, with $n \in \mathbb{Z}$. (The physical magnetic flux is $2 \pi n/e$.)  In the last step above we've written $m=\phi_0= e v$ in terms of the canonically normalized vev of $\phi$. 
 
 While this discussion describes an infinitely long string, we note that because of the Dirac 
 charge quantization condition $eg= n/2$ with $g$ in units of $4 \pi/e$, the ends of 
 the flux tube can support a Dirac monopole (or anti-monopole) having a magnetic flux of $2 \pi n/e$, such 
 that the monopole-anti-monopole pair has zero net flux. 

Outside of the core of the string the profile of the gauge field also falls exponentially, with a rate also 
set by $\phi_0$. The characteristic thickness of the string is thus of $\mathcal{O}(m^{-1})$ the gauge boson mass.

\section{WKB estimate in parabolic Coordinates} 
\label{appendix:parabolic-coordinates}

Parabolic coordinates are a useful coordinate system to work in when a Hamiltonian with spherical symmetry is perturbed by an external potential having only azimuthal symmetry. 

The Hamiltonian describing the bound states of the monopole-anti-monopole system is 
\be 
H_0 = \frac{p^{2}}{2 \mu} - \frac{\alpha_D}{r}e^{-m_{D}r}  
+ \hat{C} r
\ee
valid in the non-relativistic limit that occurs when the mass $m_D$ of the gauge boson $A_D$ is much less than the monopole mass $M$; 
$m_D \ll M $. Here $\hat{C}=C \pi v^2_D$ is the tension, and $r$ is the distance separating the monopole and anti-monopole, and the reduced mass $\mu$ of the system is $\mu=M/2$. The ground state is dominated by Coulombic interactions 
when $m_D L_{0} \ll 1$, which we assume henceforth. We therefore treat the Yukawa potential as a Coulomb interaction, setting the exponential factor to unity.  

By electromagnetic duality, in a static external magnetic field $B_{\rm eff}(d)$ the total Hamiltonian of the system is 
\be 
H=H_0 + g Q_m B(d) z ~.
\ee
In this background, the ground state is quantum mechanically unstable to decay, as the potential is unbounded from below at large negative $z$. 

In what follows we compute the decay probability in the quasi-classical approximation. In the case of a 
hydrogen-like system, this is 
a problem in Landau and Lifshitz \cite{Landau:1991wop} that happily enough comes with a solution, and can 
be easily adapted to the situation here where the Hamiltonian also has a confining potential. 

In the quasi-classical approximation the decay rate $\Gamma$ is estimated as given by 
the probability current evaluated just on the other side of the classical barrier, or second classical 
turning point, integrated over the plane perpendicular to the direction of the unstable direction. Namely, 
\be
\Gamma= \int \limits_0^\infty d \rho (2 \pi \rho) |\psi|^2 v _z 
\label{app:WKB-decay-rate}
\ee 
where $\rho$ is the radius in cylindrical coordinates and $v_z=p_z/\mu$ is the classical 
expression for the velocity in the (negative) $z$-direction. 

To proceed, following Landau and Lifshitz \cite{Landau:1991wop}, it is helpful at this point to introduce 
parabolic coordinates 
\be
\xi=r+z~,~~ \eta=r-z
\ee
where $\xi$ and $\eta$ are strictly positive because $r \geq |z|$. Thus $r=\sqrt{\eta \xi}$ and $\rho=\sqrt{\rho^2-z^2}$.
Next, we approximate the Yukawa factor $\exp \left(-m_D r \right) \approx 1$ and rescale variables from physical units given in the above equations to natural units denoted by an overline, 
\bea
E &=& \frac{\mu}{\hbar^2} \alpha_D ^2 \overline{E} \\
(\xi, \eta) &=& \frac{\hbar^2}{\alpha_D \mu }  (\overline{\xi}, \overline{\eta})\\
g Q_m B(d) &=& \alpha_D^3 \frac{\mu^2}{\hbar^4} \overline{B} \\
\hat{C}&=& \alpha_D ^3 \frac{\mu^2}{\hbar^4}  \overline{C}~.
\eea 
Then assuming the wavefunction $\psi$ is separable in parabolic coordinates (and henceforth we drop the overline on quantities in natural units)
\be 
\psi=\frac{\chi_1}{\sqrt{\xi}} \frac{\chi_2}{\sqrt{\eta}} e^{i m \phi} 
\ee 
where $m$ is the azimuthal quantum number, 
one obtains two independent one-dimensional Schr\"{o}dinger equations for 
$\chi_1$ and $\chi_2$ given by 
\bea
E_1 \chi_1&=& -\frac{1}{2} \frac{d^2}{d \xi^2} \chi_1 + U_1(\xi) \chi_1  \\
E_2 \chi_2 &=& -\frac{1}{2} \frac{d^2}{d \eta^2} \chi_2 + U_2(\eta) \chi_2 
\eea
with $E_{1,2}=E_n/4=-1/(8 n^2)$, and 
\bea 
U_1 &=& -\frac{\beta_1}{2\xi} + \frac{m^2-1}{8 \xi^2} 
+\frac{1}{8}(B+C) \xi \\
U_2 &=& -\frac{\beta_2}{2\eta} + \frac{m^2-1}{8 \eta^2} 
-\frac{1}{8}(B-C) \eta 
\eea
Here $\beta_1$ and $\beta_2$ are separation constants, 
with $\beta_1+\beta_2=1$ for the Coulomb potential. Since both $\xi \geq 0$, $\eta \geq 0$, $C>0$ and $B >0$, 
the potential in the $\xi$ direction remains confining, whereas quantum mechanical tunneling in the $\eta$ direction 
can occur, provided $B >C$ which we will assume. 
The Coulomb potential provides a barrier to tunneling, but its height is set by the small magnetic fine structure constant. 
On the other side of that barrier at the second classical turning point, the potential in the
$\eta$ direction is unbounded from below, and corresponds in physical units to large negative $z$.

For the ground state, $\beta_1=\beta_2=1/2$ and $m=0$, leading to 
\bea 
U_2&=& -\frac{1}{4 \eta}- \frac{1}{8 \eta^2}- \frac{1}{8} (B-C) \eta 
\eea

The wavefunction of the ground state is 
\bea 
\chi_1(\xi) &=&\sqrt{\xi} e^{-\xi/2} \\
\chi_2(\eta) &=&\frac{\sqrt{\eta}}{\sqrt{\pi}} e^{-\eta/2} 
\eea
In the background magnetic field $\chi_1$ remains well-approximated by the unperturbed wavefunction because the one-dimensional potential for $\chi_1$ has no unbounded from below directions. The quasi-classical 
approximation for $\chi_2$ is 
\bea 
\chi_2(\eta) &=& \frac{C}{\sqrt{p(\eta)}} \exp \left(i \int \limits^{\eta} _{\eta_0} d \eta^\prime p(\eta^\prime) + i 
\frac{\pi}{4}\right)
\eea
with 
\bea 
p &=& \sqrt{2(E-U_2)} \\
 &=& \frac{1}{2} \sqrt{B \eta -1 + \frac{2}{\eta} + \frac{1}{\eta^2}}
\eea
the classical momentum. Here $\eta_0 \gg 1$ is a point well-inside the classically forbidden region, 
further satisfying $B \eta_0 \ll 1$ so that near the first classical turning point 
$p_0 \simeq 1/2$. The constant 
\be
C=\sqrt{\frac{\eta_0 p_0}{\pi}} e^{-\eta_0/2} 
\ee
is obtained from matching the expression above for $\chi_2$ to the ground state wavefunction at $\eta=\eta_0$, giving 
\cite{Landau:1991wop}
\bea 
|\chi(\eta)|^2 &=& \frac{\eta_0 \xi}{\pi} \frac{|p_0|}{p} e^{-\xi} \exp\left(-2 \int \limits^{\eta} _{\eta_0} d \eta^\prime p(\eta^\prime) -\eta_0\right)
\eea
With $B > C$, then in the small tension and 
small mixing limit $C< B \ll 1$, the second classical turning point $\eta_{1}$ is simply 
\be
\eta_{1} = \frac{1}{B - C}
\ee
Then after expanding $p$ to next order in $1/\eta$, one finds that the explicit dependence of the wavefunction 
on the initial point $\eta_0$ is canceled by the dependence arising from the integral over $p$, so that 
to $O(B \eta_0)$ and $O(1/\eta^2)$, the quasi-classical wavefunction close to the second turning point $\eta_1$ is 
\bea 
|\chi|^2 &=& \frac{4 \xi}{B \pi} \frac{1}{\sqrt{B \eta-1}} e^{-\xi} e^{-\frac{2}{3B}} e^{O(B \eta_0)}
\eea
One then sees that the WKB approximation is justified provided that $B \ll 1$ in natural units.
Since the support of the decay probability is at $\eta \gg \xi$, one has 
$d \rho \simeq \sqrt{\eta} d \xi/(2 \sqrt{\xi})$, 
and with $v_z=\sqrt{B \eta -1}$,  integrating $|\chi|^2$ in (\ref{app:WKB-decay-rate})
the decay probability in physical units is finally \cite{Landau:1991wop}
\bea 
\Gamma &=& \left( \frac{4 \alpha^5 _D \mu^3}{g Q_m B(d)}\right) e^{-\frac{2 \alpha^3_D \mu^2}{3 g Q_m B(d)}} 
\eea 

The condition $\eta_{1} \gg 1$ is then simply 
\be 
g Q_m B(d) \ll \left(\frac{\alpha_D}{\hbar^2}\right)^3 \mu^2 
= \frac{1}{4} \alpha_D^3 M^2/ \hbar^6
\ee
in terms of physical units. This condition is equivalent to $\gamma \gg 1$ with $\gamma$ given by Eq. \ref{eq:WKBvalid}.

\section{Magnetic Plasma Oscillations}
\label{app:osc}

In Ref.~\cite{Turner:1982ag}, it was suggested that monopoles may induce plasma oscillations. These would be rapid changes in the monopole density, and the frequency of these oscillations is $\omega = \sqrt{\rho_{M} g^{2} Q_{m}^{2} /M^{2}}$. If such oscillations were occurring on short time-scales compared to the dynamo timescale, the observed B-field would be rapidly changing. Thus given that this does not occur, one needs, $\omega^{-1} \gg \tau_{{\rm dyn}}$, which implies
\be
Q_{m} \lesssim \frac{M}{g\tau_{{\rm dyn}} \sqrt{\rho_{M}}}. 
\label{eq1:osc}
\ee
In the case of ordinary monopoles, Ref.~\cite{1987ApJ...321..349P} argues that for oscillations not to be subject to damping the phase velocity {$\sim \omega/k$} of the oscillations has to exceed the galactic monopole virial velocities $v_{0} \sim$ 200 km/s. Thus in order for Landau damping not to occur on $2 \pi {k^{-1}}\sim$ kpc scales, we have a lower bound on $Q_{m}$ 
\be
Q_{m} \gtrsim \frac{M}{g \sqrt{\rho_{M}}}\left(\frac{2 \pi}{{\rm kpc}}\right)v_{0}
\label{eq2:osc}
\ee
However, Eqs.~\ref{eq1:osc} and ~\ref{eq2:osc} cannot be simultaneously satisfied, 
{simply because $2 \pi v_0 \tau_{\rm dyn}/{\rm kpc} \sim O(4 \cdot 10^2) \gg 1$}. We therefore conclude that plasma oscillations will not play a significant role.


\bibliographystyle{JHEP}
\bibliography{Ref.bib}

\end{document}